\documentclass[12pt, prd, nofootinbib,preprint,superscriptaddress]{revtex4}
\usepackage{graphicx} 
\usepackage[utf8]{inputenc}
\usepackage[T1]{fontenc}
\usepackage{comment}
\usepackage{amsmath,amssymb,ulem}
\usepackage{epsfig}
\usepackage{graphicx}
\usepackage[usenames,dvipsnames]{color}
\usepackage{subfigure}
\usepackage{slashed}
\usepackage[colorlinks,citecolor=blue]{hyperref}
\usepackage{pdfpages}
\usepackage{color}
\usepackage{lipsum}

\usepackage{tikz-feynman}
\tikzfeynmanset{compat=1.1.0}

\usepackage{tikzsymbols}

\begin{document}
\title{On the Maximal CP Violation in Leptogenesis\\  with Two Right Handed Neutrinos
}
\author{Swapnil Dutta}
\email{swd20@pitt.edu}
\affiliation{Pittsburgh Particle Physics, Astrophysics, and Cosmology Center, Department of Physics and Astronomy, University of Pittsburgh, Pittsburgh, PA 15206, USA}


\begin{abstract}
We present a generalized exact analytical expression for the upper bound on the CP-violating decay asymmetry in the case of standard vanilla leptogenesis within the minimal type-I seesaw with  two Right Handed Neutrinos (RHNs) case scenario as a function of the effective neutrino mass in the hierarchical limit. We derive the bound using a novel analytical treatment of the roots of a quartic equation arising from the conditions for successful leptogenesis, in which the CP asymmetry, effective neutrino mass, and other physical quantities enter as parameters. We find that the upper bound on CP asymmetry is a monotonically increasing function of the effective neutrino mass, asymptotically saturating to a global maximum at the large effective mass regime. We find that our exact analytical expression of the upper bound as a function of the effective neutrino mass is smaller than the previous analytical results in the literature.
We also find that the global maximum on CP asymmetry with two RHNs obtained at the large effective mass regime is less than the  Davidson-Ibarra bound  corresponding to the standard vanilla leptogenesis scenario with three RHNs, leading to a more stringent  bound on the lightest RHN mass for the scenario with two RHNs, as expected from some previous results in literature. We find that the deviation from the standard Davidson-Ibarra bound becomes most dramatic for inverted neutrino mass hierarchy, by nearly two orders of magnitude.
\end{abstract}

\maketitle
\section{Introduction}
Solar~\cite{Cleveland:1998nv, GALLEX:1998kcz, SAGE:1999nng,  Super-Kamiokande:2001ljr,Smy:2002rz,SNO:2001kpb, SNO:2002tuh} and atmospheric~\cite{Super-Kamiokande:1998kpq, Super-Kamiokande:1998uiq, Super-Kamiokande:2000ywb, Soudan-2:1999jbo, Alberico:2003kd} neutrino observations, and later reactor neutrino experiments~\cite{KamLAND:2002uet, DayaBay:2013yxg, RENO:2012mkc, DoubleChooz:2011ymz, DoubleChooz:2014ppz} and accelerator neutrino experiments~\cite{K2K:2001nlz,K2K:2002icj,K2K:2004iot,MINOS:2006foh,T2K:2011ypd,NOvA:2016kwd}, have shown evidences that neutrinos undergo flavor oscillations~\cite{Cleveland:1998nv, GALLEX:1998kcz, SAGE:1999nng,  Super-Kamiokande:2001ljr, Smy:2002rz, SNO:2001kpb,SNO:2002tuh, Super-Kamiokande:1998kpq, Super-Kamiokande:1998uiq, Super-Kamiokande:2000ywb, Soudan-2:1999jbo, Alberico:2003kd,KamLAND:2002uet, DayaBay:2013yxg, RENO:2012mkc, DoubleChooz:2011ymz, DoubleChooz:2014ppz, K2K:2001nlz,K2K:2002icj,K2K:2004iot,MINOS:2006foh,T2K:2011ypd,NOvA:2016kwd} which show that neutrinos are not massless~\cite{Maki:1962mu, Pontecorvo:1967fh}, contrary to the prior perception. Since then, numerous hypotheses have been proposed to explain the origin of neutrino masses. 
One of the most elegant, economical and plausible frameworks among them is the type-I seesaw~\cite{gellmann2013complexspinorsunifiedtheories, PhysRevLett.44.912, Yanagida:1979as, Minkowski:1977sc, GellMann:1980vs,Glashow:1979nm,Mohapatra:1979ia,Schechter:1980gr}.
In addition to explaining the origin of neutrino mass, the type-I seesaw can also explain the origin of the matter-antimatter asymmetry through a mechanism called leptogenesis~\cite{Fukugita:1986hr}.

In order to explain the observed masses of the active neutrinos and to be compatible with leptogenesis, a minimum of two RHNs are required. Various aspects of the minimal scenario have been explored in the literature; see, e.g., Refs~\cite{King:1999mb, Frampton:2002qc, 2002PhLB..548..204A, Appelquist:2003uu, 2004EPJC...32..235R, King:2002nf,Dreiner:2003yr, Ibarra:2003up, Ibarra:2005qi, Bhattacharya:2006aw,Antusch:2011nz, Li:2017zmk,Xing:2020ald,Okada:2025daq,Datta:2025vyu,Ghoshal:2025iil,Spalding:2026jia,King:2025eqv,Zhang:2024weq,Guo:2006qa,DiBari:2005st}. 
In this paper, we revisit the conditions for leptogenesis in the minimal two RHN seesaw model, focussing on the simplest case of vanilla leptogenesis. 
As we will show in detail, the requirement of successful leptogenesis fixes some of the otherwise undetermined parameters of the two RHN type-I seesaw in terms of only two unknown quantities: the effective neutrino mass and the mass of the lightest RHN. Together, these two parameters completely determine the final lepton and baryon asymmetries.
And mathematical analysis of these deterministic relations will lead to generalized exact analytical expressions of new bounds on parameters related to the standard vanilla leptogenesis in the context of two RHN scenario in the hierarchical limit $m_{N_1}\ll m_{N_2}$, where $m_{N_1}$ and $m_{N_2}$ are the masses of the lighter and the heavier RHN respectively. The most significant of these bounds is the bound on CP violation generated from the decay of the lighter RHN $N_1$ that generates the lepton asymmetry. We find that the general exact analytic expression of the upper bound on CP violation that we derive is lower than the analytical upper bounds shown in \cite{Ibarra:2003up,Guo:2006qa,DiBari:2005st}.
Also, numerical scans have been done to obtain the possible range of CP violation and the mass of the lightest RHN in two RHN scenario compatible with leptogenesis and numerically  bounds have been obtained on them~\cite{Bhattacharya:2006aw,Xing:2020ald,Okada:2025daq}. We, however, for the first time to our knowledge, derive the generalized exact analytical expression of the upper bound on CP violation explicitly as a function of effective neutrino mass $\tilde{m}_1$ for standard vanilla leptogenesis in 2 RHN scenario in the strong hierarchical limit $m_{N_1}\ll m_{N_2}$. Also, we use a novel mathematical technique of analyzing the roots of a quartic equation to derive the analytical expressions of the CP asymmetry. This novel mathematical technique, we hope, will in future provide new insights and probes to the community to further explore the theoretical aspects of leptogenesis in the two RHN scenario. 

We arrange the contents of this paper in the following way. In section \ref{sec:The 2 RHN seesaw I model}, we introduce the type-I seesaw model in the context of 2 RHN scenario. In section \ref{sec: Vanilla leptogenesis in the 2 RHN model}, we elaborate the standard vanilla leptogenesis mechanism in the context of the 2 RHN type-I seesaw. We show how the parameters of the 2 RHN type-I seesaw can be related to the quantities that determine leptogenesis. As mentioned in the previous paragraph, we will see that leptogenesis conditions determine the parameters of the 2 RHN type-I seesaw except the mass of heavier RHN $m_{N_2}$ up to $\pm$ signs in terms of just two parameters that control the final lepton asymmetry as they are related via some equations. In this context, we will see that from those equations, we can obtain a quartic equation in terms of one of the unknown parameter and this equation will play a major role in the subsequent sections. In section \ref{sec: A new upper bound in CP violation}, we will see how the requirement of physically acceptable solutions of the quartic equation implies the existence of an upper bound of the CP asymmetry required for leptogenesis for both normal and inverted hierarchy. We believe that this is the exact analytical expression of the effective neutrino mass dependent upper bound on CP-violating decay asymmetry. There are numerical differences between our bound and  analytical bounds presented earlier in literature ~\cite{Ibarra:2003up,Guo:2006qa,DiBari:2005st} due to various approximations made there and our exact analytical expression gives stronger bound\footnote{Also, an analytical upper bound of CP violation for the 2RHN scenario was derived in \cite{Drees:2024hok} but it was a more general global upper bound and was not dependent on $\tilde{m}_1$. As we will see in section \ref{sec: A new upper bound in CP violation}, the global upper bound is reached in large $\tilde{m}_1$ limit and our global upper bound agrees with the one derived in \cite{Drees:2024hok}.}.
We provide the key arguments and the line of reasoning in this section without elaborating on the mathematical details for the convenience of the readers, which will be elaborated in the appendices. In section \ref{sec: A new lower bound on lightest RHN mass}, we derive a lower bound on the mass of the lightest RHN that is compatible with leptogenesis using the new upper bound in CP violation that we derived in the previous section. Finally, we summarize and conclude the main article with section \ref{sec: Conclusions} following which there would be appendices mentioned earlier. In Appendix \ref{sec:Nature_of_Roots}, we elaborate on the nature of roots of a general quartic equation, followed by Appendix \ref{app:Nature_of_Roots_Cubic_Equation}, where we do the same for cubic equations. In appendix \ref{app:RuleofSigns}, we briefly elaborate the Descartes' rule of signs, which is used to find the sign of real roots of a general polynomial equation, along with an example showing its application. Finally, in Appendix \ref{app:DetailedAnalysis}, we elaborate  all the mathematical analyzes and details regarding the nature of roots in the quartic equation corresponding to the 2 RHN scenario which support and complete the arguments and the line of reasoning presented in section \ref{sec: A new upper bound in CP violation}. We only present the detailed arguments taking the case of the normal hierarchy of neutrino mass ordering although they still remain valid for inverted hierarchy.

\section{The 2 RHN type-I seesaw model}\label{sec:The 2 RHN seesaw I model}

In the standard type-I seesaw mechanism with 2 RHNs, the Lagrangian of the lepton sector contains the following interactions:
\begin{equation}
     \label{eq:SeesawI}
    -\mathcal{L}\supset \overline L y N_R H +\frac{1}{2} \overline N_{\! R}^{\,c} M_R N_R  + {\rm h.c.} \, 
\end{equation}
where $N_R$ denotes the two generations of heavy RHNs 
, $L$ denotes the three generations of $SU(2)_L$ lepton doublets of the Standard Model (SM), 
$H$ is the SM Higgs doublet, $y$ denotes the Yukawa coupling,
which in the case of 2 RHNs is a $3\times 2$ matrix, and $M_R$ denotes the mass matrix of the RHNs, which in this case is a $2\times 2$ matrix. We can move to a basis where $M_R$ is diagonal and real by applying suitable flavor transformations to the  two RHNs. 
In this basis, the Yukawa matrix has 6 complex elements, which amount to six moduli and six phases, and there are also the two real, positive RHN Majorana masses contained in $M_R$.
This amounts in total to eight moduli and six phases. However, phase rotations of the three generations of $SU(2)_L$ of lepton doublets render three of the six phases of the Yukawa matrix unphysical. Therefore, in this convenient flavor basis, the number of physical parameters amount to eight moduli and three phases. 

After the electroweak symmetry breaking, the Higgs field gets a vacuum expectation value (vev) so that we have $\langle H\rangle={(v/\sqrt{2},0)}^T$. Then, the SM neutrinos acquire mass terms and the interactions in the Lagrangian which give masses to the neutral leptons (the three active neutrinos of the SM and the 2 RHNs), become
\begin{align}
\label{eq:neutrino-mass}
 -{\cal L_{\textrm{mass}}} &\supset \overline \nu_L M_D N_R + \frac{1}{2} \overline N_{\! R}^{\,c} M_R N_R  + {\rm h.c.} \,\\
 &= \frac{1}{2} \biggl(\overline \nu_L ~~ \overline N_{\! R}^{\,c} \biggr)
 \left( 
  \begin{array}{cc}
  0 & M_D \\
  M_D^T & M_R
  \end{array}
 \right)
 \left(  
 \begin{array}{c}
 \nu_L^c  \\
 N_R 
 \end{array}
 \right) +{\rm h.c.}  \\
& = \frac{1}{2} \, \overline n_{\! R}^{\,c} \, V^T
\left(
 \begin{array}{cc}
 0 & M_D \\
 M_D^T & M_R
 \end{array}
\right)
 V n_R+{\rm h.c.} 
\nonumber \\
 & = \frac{1}{2} \, \overline n_{\! R}^{\,c} \, M_n \, n_R + {\rm h.c.} \, ,\label{eq:Diagnolazed mass term}
 \end{align}
where $M_D=yv/\sqrt{2}$ and $M_R$ (already defined in Eq.~\eqref{eq:SeesawI}) are the Dirac and Majorana mass matrices respectively and $M_n=\textrm{diag}(m_{\nu_1}=0,m_{\nu_2},m_{\nu_3},m_{N_1},m_{N_2})$ is the diagonalized neutrino mass matrix with mass eigenvalues arranged in increasing order from $m_{\nu_1}$ to $m_{N_2}$ assuming normal neutrino mass ordering while $V$ is an unitary transformation relating the flavor basis $(\nu_L^c ~ N_R )^T$  with the mass basis $n_R$ through $(\nu_L^c ~ N_R )^T =  V n_R$. For inverted neutrino mass ordering, we have $M_n=\textrm{diag}(m_{\nu_3}=0,m_{\nu_1},m_{\nu_2},m_{N_1},m_{N_2})$ with mass eigenvalues arranged in increasing order from $m_{\nu_3}$ to $m_{N_2}$. From now on, we will implicitly assume normal ordering unless explicitly stated otherwise.  All the results, mathematical steps and reasonings will remain equally valid for inverted ordering by relabelling the light neutrino mass eigenstates as $m_{\nu_1}\rightarrow m_{\nu_3}$, $m_{\nu_2}\rightarrow m_{\nu_1}$ and $m_{\nu_3}\rightarrow m_{\nu_2}$.

It is to be noted that in the 2 RHN scenario, lightest active neutrino is strictly massless.
Therefore, $n_R$ contains the five  neutrino mass eigenstates, corresponding to one massless neutrino along with four massive Majorana neutrinos, as seen from the form of the mass term in Eq.~\eqref{eq:Diagnolazed mass term}. The matrix $V$, which is a $5\times 5$ matrix, can be expressed as
\begin{align}
\label{eq:V}
V 
& \simeq
\left(
 \begin{array}{cc}
 U^{*} & -i U^{*} \sqrt{d_{l}} R^{\dag} \sqrt{d^{-1}_{h}} \\
 -i \sqrt{d^{-1}_{h}} R \sqrt{d_{l}} & 1
 \end{array}
\right),
\end{align}
where $U$ is the Pontecorvo–Maki–Nakagawa–Sakata (PMNS) matrix~\cite{Maki:1962mu,Pontecorvo:1967fh}, $d_{l}$ ($d_{h}$) is the diagonal mass matrix of the light, mostly active (heavy, mostly sterile) neutrinos, and $R$ is a $2\times 3$ complex orthogonal matrix given by~\cite{Ibarra:2003xp,Ibarra:2003up} \begin{align}
    \label{eq:RT}
     R = \begin{pmatrix}
 0& \cos(z) & \pm \sin(z)\\
 0& -\sin(z) & \pm \cos(z)
 \end{pmatrix} \hspace{2pt}.
\end{align}
The Yukawa matrix $y$ can be expressed in terms of the $R$ matrix and the other matrices that appear above in Eq.~\eqref{eq:V} as
\begin{align}
    \label{eq:CIS}
    y = \frac{i \sqrt{2}}{v}  U \sqrt{d_{l}} R^{T} \sqrt{d_{h}} \, .
\end{align}
This is the well known Casas-Ibarra parametrization~\cite{Casas:2001sr}.

The PMNS matrix $U$ can be parameterized in terms of three mixing angles, one Dirac CP violation phase and one majorana phase, all of which are in principle possible to be measured in experiments. This amounts to three moduli and two phases which are known in principle. Also, we have two active neutrino masses $m_{\nu_2}$ and $m_{\nu_3}$, which are completely determined from the solar and atmospheric neutrino parameters~\cite{Super-Kamiokande:2001ljr,Smy:2002rz,SNO:2001kpb, SNO:2002tuh,Super-Kamiokande:2000ywb,Alberico:2003kd,KamLAND:2002uet} $\Delta m_{\textrm{sol}}^2$ and $\Delta m_{\textrm{atm}}^2$ respectively.
So, these add up to a total of five moduli and two phases below the electroweak scale. These parameters, in principle, can be experimentally measured. So, out of the 8 moduli and 3 phases of the type-I seesaw model at energies above the electroweak symmetry breaking, since at low energies, 5 moduli and two phases can be detected, it implies that there are three moduli and one phase that cannot be measured. Out of them, two of the moduli are the masses of the RHNs $m_{N_1}$ and $m_{N_2}$. The remaining two parameters, one moduli and one phase, are contained in the complex orthogonal $R$ matrix introduced in Eq.~\eqref{eq:V}. 

We will show that the requirement of successful leptogenesis fixes these two remaining parameters by completely determining the $R$ matrix in terms of parameters which determine the final baryon asymmetry, namely $m_{N_1}$ and the \textit{effective neutrino mass} $\tilde{m}_1$, which will be introduced in section \ref{sec: Vanilla leptogenesis in the 2 RHN model} . The information about these unknown parameters are lost as the RHNs decouple at low energies. 


\section{Vanilla leptogenesis in the 2 RHN model}\label{sec: Vanilla leptogenesis in the 2 RHN model}

In the hierarchical limit $m_{N_1}\ll m_{N_2}$, the CP-violating and the lepton number-violating decays of the lighter RHN $N_1$ produces a lepton asymmetry, which gets converted to a baryon asymmetry by the electroweak sphalerons. This is the standard vanilla leptogenesis in the context of the scenario with two RHNs. The baryon-to-photon ratio in vanilla 
leptogenesis can be parameterized as~\cite{Buchmuller:2004nz}
\begin{align}
    \label{eq:YB}
    \frac{n_{B}}{n_{\gamma}} &=\frac{3}{4}\frac{a_{\textrm{sph}}}{f} |\varepsilon_1| \kappa_f \simeq 0.96 \times 10^{-2} \, |\varepsilon_{1}| \, \kappa_{f} \, ,
\end{align}
where $a_{\textrm{sph}}=\frac{28}{79}$ is the fraction of $B-L$ asymmetry that is converted to baryon asymmetry via sphaleron process, $f = \frac{2387}{86}$ is the entropy dilution factor from the onset of leptogenesis till recombination, 
$\kappa_f$ is the final efficiency factor that takes into account the effect of $B-L$ washout from different processes  
and $\varepsilon_1$ is the decay CP asymmetry generated from the decay of $N_1$, which, in the hierarchical limit, can be written as~\cite{Covi:1996wh,Davidson:2002qv}
\begin{align}
    \label{eq:epsilon}
    \varepsilon_{1} & \simeq -\frac{3}{16 \pi (y^{\dagger}y)_{11}}  {\rm Im}\left[\left(y^{\dagger} y\right)^2_{12}\right] \frac{m_{N_{1}}}{m_{N_{2}}} 
    = \frac{3}{8\pi} \frac{m_{N_1}}{v^2} \frac{\sum_i m^2_{\nu_i} {\rm Im}(R_{1i}^2)}{\sum_i m_{\nu_i} |R_{1i}|^2 }.
\end{align}
With the structure of the $R$-matrix given in Eq.~\eqref{eq:RT}, $\varepsilon_1$ becomes
\begin{equation}
    \label{eq:epsilon2RHN}
    \varepsilon_1 =  \frac{3 m_{N_1} \left({m_{\nu_3}^2}-{m_{\nu_2}^2}\right)}{16\pi v^2 \tilde{m}_1} \textrm{Im}\left(\sin^2{z}\right) \hspace{2pt},
\end{equation}
where~\cite{Buchmuller:2004nz}
\begin{equation}
\label{eq:EffNeutrinoMass}
    \tilde{m}_{1} \equiv \frac{v^{2}~\left(y^{\dag} y\right)_{11}}{2 m_{N_{1}}} = m_{\nu_1} |R_{11}|^{2} + m_{\nu_2} |R_{12}|^{2} + m_{\nu_3} |R_{13}|^{2}
\end{equation}
is the \textit{effective neutrino mass} (introduced earlier in section \ref{sec:The 2 RHN seesaw I model}), which for the structure of the \textit{R}-matrix given in Eq.~\eqref{eq:RT}, becomes
\begin{equation}
\label{eq:m1tilde2RHN}
    \tilde{m}_1= m_{\nu_2}{|R_{12}|}^2 + m_{\nu_3}{|R_{13}|}^2 = m_{\nu_2} |\cos^2(z)| + m_{\nu_3} |\sin^2(z)|\hspace{2pt}.
\end{equation}
As is evident from Eq.~\eqref{eq:m1tilde2RHN}, there is a theoretical lower bound on $\tilde{m}_1$ in the 2 RHN scenario given by
\begin{equation}
    \label{eq:effectiveneutrinomasslowerbound}
    \tilde{m}_1> m_{\nu_2}(|\cos^2(z)|+|\sin^2(z)|)\geq m_{\nu_2}|\cos^2(z)+\sin^2(z)|=m_{\nu_2}
\end{equation}
where we used the inequality $m_{\nu_3}>m_{\nu_2}$  and  Eq.~\eqref{eq:m1tilde2RHN}. 


The the efficiency factor $\kappa$ tracks the production and washout of $B-L$ asymmetry during leptogenesis. In general, it is a function of the evolution variable $y\equiv m_{N_1}/T$, which increases as the universe cools, the effective neutrino mass $\tilde{m}_1$ defined in Eq.~\eqref{eq:EffNeutrinoMass}, the mass $m_{N_1}$ of the lightest RHN, and the quadratic sum of neutrino masses $\Bar{m}\equiv\sqrt{m_{\nu_1}^2+m_{\nu_2}^2+m_{\nu_3}^2}$~\cite{Barbieri:1999ma,Buchmuller:2002rq}. The final efficiency factor appearing in  Eq.~\eqref{eq:YB} is conventionally defined as $\kappa_f\equiv\kappa(y\rightarrow\infty,\tilde{m}_1, m_{N_1},\bar{m})$~\cite{Buchmuller:2002rq}. 
For two RHNs, since the masses of the active neutrinos are known, their quadratic sum $\Bar{m}$ is determined. Therefore, $\kappa_f $ is effectively governed by just two independent parameters $m_{N_1}$ and $\tilde{m}_1$. Also, from Eq.~\eqref{eq:epsilon2RHN}, it is evident that the CP asymmetry $\varepsilon_1$ is also a function of these two independent parameters since ${m_{\nu_3}^2}-{m_{\nu_2}^2}$ has been experimentally measured. Therefore, it follows from Eq.~\eqref{eq:YB} that the lepton asymmetry, and hence, the baryon asymmetry, generated by leptogenesis can be parameterized by $m_{N_1}$ and $\tilde{m}_1$.

The observed baryon asymmetry of the universe is given by $n_{B}/n_{\gamma}\vert_{\rm obs} = 6 \times 10^{-10}$~\cite{Planck:2018vyg,Yeh:2026pil}. Comparing this with Eq.~\eqref{eq:YB} gives 
\begin{equation}
    \label{eq:CP violationObservedBaryonAsymmetry}
    |\varepsilon_1|
    \simeq \frac{6.2\times10^{-8}}{\kappa_f}
\end{equation}
which is the CP violation that is required to generate the observed baryon asymmetry. This should be equal to $|\varepsilon_1|$ c.f. Eq.~\eqref{eq:epsilon2RHN} for leptogenesis in 2 RHN scenario to be successful.
 

For $m_{N_1}\ll10^{14}$ GeV, the final efficiency factor $\kappa_f$  depends, to a good approximation, only on $\tilde{m}_1$~\cite{Buchmuller:2002rq,Giudice:2003jh}. Through Eq.~\eqref{eq:CP violationObservedBaryonAsymmetry}, the CP asymmetry required to reproduce the observed baryon asymmetry therefore depends predominantly on $\tilde{m}_1$ as well. Semi-analytical expressions for $\kappa_f(\tilde{m}_1)$ have been obtained by fitting to the numerical solutions of the Boltzmann equations for leptogenesis~\cite{Buchmuller:2002rq,Giudice:2003jh}. Assuming zero initial abundance
of $N_1$, we will use the semi-analytical form for zero initial abundance~\cite{Giudice:2003jh}
\begin{equation}
    \label{eq:kappaonm1tilde}
    \kappa_f(\tilde{m}_1)\approx {\left[\frac{3.3\times10^{-3}\textrm{ eV}}{\tilde{m}_1}+{\left(\frac{\tilde{m}_1}{0.55\times10^{-3}\textrm{ eV}}\right)}^{1.16}\right]}^{-1} \hspace{5pt}.
\end{equation}
The results in \cite{Giudice:2003jh}  account for the finite temperature thermal corrections and a proper treatment of real intermediate state contributions to the $\Delta L=2$ scattering processes.
Incorporating Eq.~\eqref{eq:kappaonm1tilde} in Eq.~\eqref{eq:CP violationObservedBaryonAsymmetry} shows that in the regime $m_{N_1}\ll10^{14}$ GeV,, the CP asymmetry $\varepsilon_1$ needed to explain the observed baryon asymmetry is exclusively dependent only on $\tilde{m}_1$. 

We note that the explicit semi-analytical functional fit to numerical solutions for thermal and dominant initial abundances of $N_1$ have not been presented in \cite{Giudice:2003jh}. Therefore, for simplicity, for results that explicitly depend on $\kappa_f$, i.e., the lower bound on $m_{N_1}$ required for successful leptogenesis (further elaborated in section \ref{sec: A new lower bound on lightest RHN mass}), we will confine ourselves to the zero initial abundance case. But the extension to thermal and dominant initial abundance cases are straight forward. Also, we emphasize that our main result, which is the exact analytical upper bound on the CP asymmetry as a function of the {effective neutrino mass} (see section \ref{sec: A new upper bound in CP violation}), does not depend on the initial abundance condition of $N_1$.

\section{An exact analytical upper bound on $\varepsilon_1$}\label{sec: A new upper bound in CP violation}
We saw in section \ref{sec: Vanilla leptogenesis in the 2 RHN model} how the final lepton (and baryon) asymmetry in vanilla leptogenesis in the strong hierarchical limit depends in general on just two parameters $m_{N_1}$ and $\tilde{m}_1$. 
In this section, we will see how constraints from leptogenesis can exactly determine the R-matrix given by Eq.~\eqref{eq:RT} at every point in this $\tilde{m}_1-m_{N_1}$ parameter space for the two RHN scenario and how algebraic analysis of these deterministic relations give a generalized exact analytical expression for the upper bound on the CP asymmetry $|\varepsilon_1|$ as a function of $\tilde{m}_1$. 
For brevity and readability, this section only presents the main mathematical steps and the underlying reasoning. Further details of the mathematical analysis are provided in the appendices, particularly Appendix \ref{app:DetailedAnalysis}. 


It is also worthwhile mentioning that the \textit{R}-matrix for the  three RHN scenario has larger number of parameters and therefore, is not possible to be uniquely or analytically determined at every point in the $\tilde{m}_1-m_{N_1}$ leptogenesis parameter space. For every $\left(\tilde{m}_1,m_{N_1}\right)$, there will be range of \textit{R}-matrices compatible with leptogenesis in the three RHN scenario. 

For algebraic convenience, we will use the parameterization $\cos^2 (z) = re^{i\theta}$ for the R matrix in Eq.~\eqref{eq:RT}. Then, in this parameterization, Eq.~\eqref{eq:epsilon2RHN} and Eq.~\eqref{eq:m1tilde2RHN} become 
\begin{equation}
    \label{eq:epsilon2RHNreparameterization}
    \varepsilon_1 =  -\frac{3 m_{N_1} \left({m_{\nu_3}^2} - {m_{\nu_2}^2} \right)r \sin{\theta}}{16 \pi v^2 \tilde{m}_1} 
\end{equation}
and
\begin{equation}
    \label{eq:m1tilde2RHNreparameterization}
    \tilde{m}_1 = m_{\nu_2} |re^{i\theta}| + m_{\nu_3} |1-re^{i\theta}|= m_{\nu_2} r + m_{\nu_3} \sqrt{1-2r\cos{\theta}+r^2} 
\end{equation}
respectively. Eliminating $\theta$ from the above equations, Eq.~\eqref{eq:epsilon2RHNreparameterization} and Eq.~\eqref{eq:m1tilde2RHNreparameterization}, 
we obtain a quartic equation in $r$, 


\begin{equation}
\label{eq:GeneralQuarticinR}
    \begin{split}
        &\frac{1}{4}{\left(\frac{{m_{\nu_2}^2}}{{m_{\nu_3}^2}}-1\right)}^2 r^4 + \frac{\tilde{m}_1m_{\nu_2}}{{m_{\nu_3}^2}}\left(1-\frac{{m_{\nu_2}^2}}{{m_{\nu_3}^2}}\right) r^3 -\frac{1}{2}\left\{1+\frac{{m_{\nu_2}^2}}{{m_{\nu_3}^2}} + \frac{{\tilde{m}_1^2}}{{m_{\nu_3}^2}} \left(1- 3 \frac{{m_{\nu_2}^2}}{{m_{\nu_3}^2}}\right) \right\} r^2\\
        +& \frac{\tilde{m}_1 m_{\nu_2}}{{m_{\nu_3}^2}}\left(1-\frac{{\tilde{m}_1^2}}{{m_{\nu_3}^2}}\right)r +\frac{1}{4}{\left(1-\frac{{\tilde{m}_1^2}}{{m_{\nu_3}^2}}\right)}^2 + k=0 \hspace{5pt},
    \end{split}
\end{equation}
where
\begin{equation}
    \label{eq:DefinitionK}
    k\equiv {\left(\frac{16\pi v^2 \tilde{m}_1 |\varepsilon_1|}{3 m_{N_1}\left({m_{\nu_3}^2} - {m_{\nu_2}^2} \right)}\right)}^2 \hspace{2pt}.
\end{equation}
Solving Eq.~\eqref{eq:GeneralQuarticinR}, we obtain $r$ as a function of five parameters $m_{\nu_2}^2$, $m_{\nu_3}^2$, $|\varepsilon_1|$, $\tilde{m}_1$ and $m_{N_1}$, from which we can obtain $\sin{\theta}$ or $\cos{\theta}$ from Eq.~\eqref{eq:epsilon2RHNreparameterization} or Eq.~\eqref{eq:m1tilde2RHNreparameterization} respectively up to a $\pm$ sign. Therefore, as seen from Eq.~\eqref{eq:RT}, we can obtain the $R$ matrix upto a $\pm$ sign.
We further note that $r$ must be real and positive as it is 
a modulus. 
Also,
it should be noted that once the neutrino mass ordering is fixed, $m_{\nu_2}^2$ and $m_{\nu_3}^2$ are determined in the two RHN scenario. 
So, for a given mass ordering, the above equation Eq.~\eqref{eq:GeneralQuarticinR} depends on three unknown parameters- $\tilde{m}_1$, $m_{N_1}$ and $|\varepsilon_1|$.
For the $R$-matrix to be physically admissible and compatible with leptogenesis, the parameters must satisfy conditions ensuring that Eq.~\eqref{eq:GeneralQuarticinR} admits at least one positive real solution, $r>0$.
These conditions are derived by analyzing the quartic equation using established results from the theory of quartic polynomials. 
As we will show, these conditions impose an exact analytical upper bound on $|\varepsilon_1|$ as a function of $\tilde{m}_1$, as aniticipated in the preceeding sections.
We reiterate that viable leptogenesis requires $|\varepsilon_1|$ above in Eq.~\eqref{eq:DefinitionK} to satisfy Eq.~\eqref{eq:CP violationObservedBaryonAsymmetry}.

The nature of the roots of the quartic equation Eq.~\eqref{eq:GeneralQuarticinR} is determined by three quantities- the discriminant $\Delta$, and two other parameters called $P$ and $D$. For a general quartic equation, the definitions of these quantities and how they determine the nature of roots are elaborated upon in Appendix \ref{sec:Nature_of_Roots}. Briefly, if $\Delta<0$, then the quartic equation has two real and two non-real roots while for $\Delta>0$, either all four roots are real or all four roots are non-real. In the latter case,  which of the two possibilities holds is further determined by some conditions on $P$ and $D$.

It turns out that for the quartic equation Eq.~\eqref{eq:GeneralQuarticinR} , the discriminant $\Delta$ is a cubic polynomial in $k$ (see Eq.~\eqref{eq:DiscriminantQuarticinR}) and therefore, the sign of $\Delta$ as a function of $k$ is determined by the roots of this cubic polynomial in $k$. We shall use the notation $\Delta(k)$ whenever we want to emphasize $\Delta$ as a cubic polynomial in $k$. The nature of the roots of this cubic polynomial $\Delta(k)$ is further determined by the discriminant $\tilde{\Delta}$ of this cubic polynomial. The dependence of the nature of roots of a cubic equation to its discriminant $\tilde{\Delta}$ is discussed in detail in Appendix \ref{app:Nature_of_Roots_Cubic_Equation}. It turns out that the discriminant $\tilde{\Delta}$ of this cubic polynomial is always positive which implies that its three roots are real and distinct. Further, using Descartes' rule of signs, elaborated upon in Appendix \ref{app:RuleofSigns}, we determined that out of these three real distinct  roots, it is always the case that two of them are positive and one is negative. Let those roots be labeled as $k_1,k_2$ and $k_3$ such that $k_1>k_2>0>k_3$. Since the coefficient of $k^3$ in $\Delta(k)>0$, we have
$\Delta(k)>0$ for $k_3<k<k_2$ and $k>k_1$, while  $\Delta(k)<0$ for $k<k_3$ and $k_2<k<k_1$. However, from Eq.~\eqref{eq:DefinitionK}, it is evident that $k\geq0$ as it is square of a real number. Therefore, in the admissible range of $k$, we have $\Delta>0$ for $0\leq k<k_2$ and $k>k_1$ while $\Delta(k)<0$ for $k_2<k<k_1$.

Upon evaluating the $P$ and the $D$ parameters in the range of $k$ where $\Delta(k)>0$, it turns out that in the region $0\leq k < k_2$, all the four solutions of the quartic equation Eq.~\eqref{eq:GeneralQuarticinR} are real and distinct while for $k>k_1$, all four solutions are non-real. Further, using Descartes' rule of signs, we can conclude that in the region $0\leq k < k_2$, out of the four real solutions for $r$, two are positive and two are negative. The two positive solutions are physically admissible and compatible with leptogenesis. In the region $k_2<k<k_1$, two roots of Eq.~\eqref{eq:GeneralQuarticinR} are real while the other two are non-real. And it can be shown by analying the roots of the derivative of the quartic polynomial in Eq.~\eqref{eq:GeneralQuarticinR} that both the real roots are negative. Therefore, the interval of $k$ where we get solutions of Eq.~\eqref{eq:GeneralQuarticinR} that are physically feasible and compatible with vanilla leptogenesis is $0\leq k<k_2$. For further details of this analysis, we refer the reader to Appendix \ref{app:DetailedAnalysis}.

Therefore, to be compatible with leptogenesis, we have the condition
\begin{equation}
\label{eq:permissiblerangeofk}
0\leq k\equiv {\left\{\frac{16\pi v^2 \tilde{m}_1 |\varepsilon_1|}{3m_{N_1} \left({m_{\nu_3}^2}-{m_{\nu_2}^2}\right)}\right\}}^2< k_2
\end{equation}
which implies 
\begin{equation}
    \label{eq:rangeofepsilon1}
     |\varepsilon_1|<\varepsilon_{\textrm{DI}} \left(1-\frac{m_{\nu_2}}{m_{\nu_3}}\right)\left(1+\frac{m_{\nu_2}}{m_{\nu_3}}\right)\left(\frac{m_{\nu_3}}{\tilde{m}_1}\right)\sqrt{k_2 (\tilde{m}_1)} \hspace{5pt},
\end{equation}
where 
\begin{equation}
    \label{eq:DIbound}
    \varepsilon_{\textrm{DI}}\equiv \frac{3 m_{N_1} m_{\nu_3}}{16\pi v^2}
\end{equation}
is the Davidson-Ibarra bound~\cite{Davidson:2002qv}. Hence, we have derived a general exact analytical expression on the upper bound of $|\varepsilon_1|$ given by
\begin{equation}
    \label{eq:ExactAnalytocalboundonepsilon1}|\varepsilon_1^{\textrm{max}}|\equiv \varepsilon_{\textrm{DI}} \left(1-\frac{m_{\nu_2}}{m_{\nu_3}}\right)\left(1+\frac{m_{\nu_2}}{m_{\nu_3}}\right)\left(\frac{m_{\nu_3}}{\tilde{m}_1}\right)\sqrt{k_2 (\tilde{m}_1)} \hspace{2pt}.
\end{equation}
We write $k_2=k_2(\tilde{m_1})$ to emphasize that the coefficients of the cubic polynomial $\Delta(k)$, and hence, its roots $k_{1,2,3}$, depend on $\tilde{m}_1$. Therefore, we see from Eq.~\eqref{eq:ExactAnalytocalboundonepsilon1} that this upper bound is an explicit function of $\tilde{m}_1$. We shall show in the subsequent subsections \ref{subsec:CPviolationnormalordering} and \ref{subsec:CPviolationInvertedOrdering} that a global maximum of $|\varepsilon_1^{\textrm{max}}(\tilde{m_1})|$, which is also the global upper bound on $|\varepsilon_1|$ in general, is reached asymptotically in the $\tilde{m}_1\rightarrow\infty$ limit.

The upper bound on $|\varepsilon_1|$ in Eq.~\eqref{eq:ExactAnalytocalboundonepsilon1} is  more restrictive than the approximate analytic upper bound presented in \cite{Ibarra:2003up}, where $\mathcal{O}\left(\frac{m_{\nu_2}}{m_{\nu_3}}\right)$ terms were neglected. It is also more restrictive than the analytical bounds presented in \cite{Guo:2006qa,DiBari:2005st}. 
We examine the dependence of $|\varepsilon_1^{\textrm{max}}(\tilde{m_1})|$ on $\tilde{m}_1$ and present some other interesting related observations  for both normal and inverted hierarchies in subsections \ref{subsec:CPviolationnormalordering} and \ref{subsec:CPviolationInvertedOrdering} respectively.

\subsection{Normal hierarchy}\label{subsec:CPviolationnormalordering}
\begin{figure}[t]
\begin{center}
\includegraphics[width=0.495\linewidth]{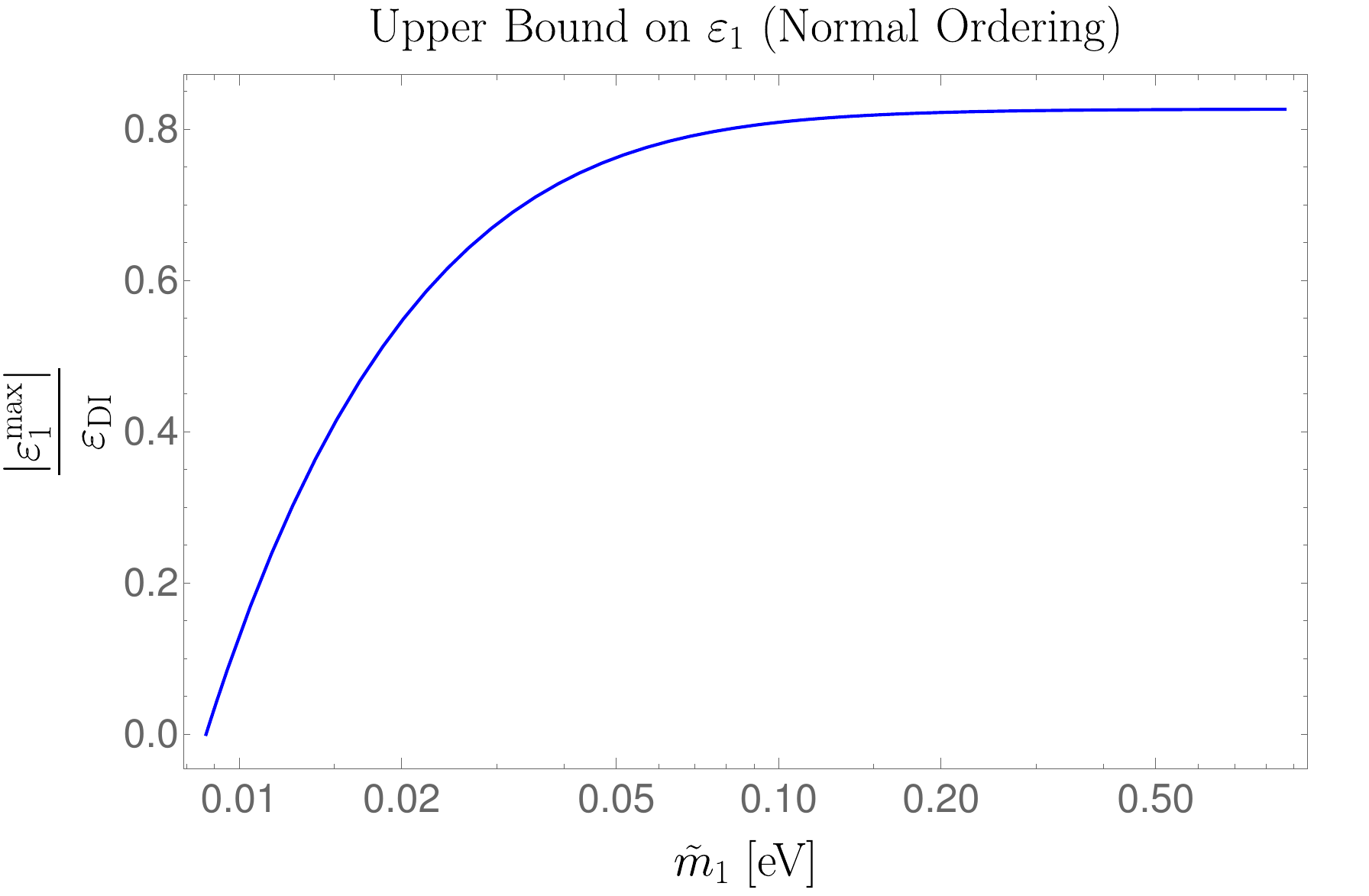}
\includegraphics[width=0.495\linewidth]{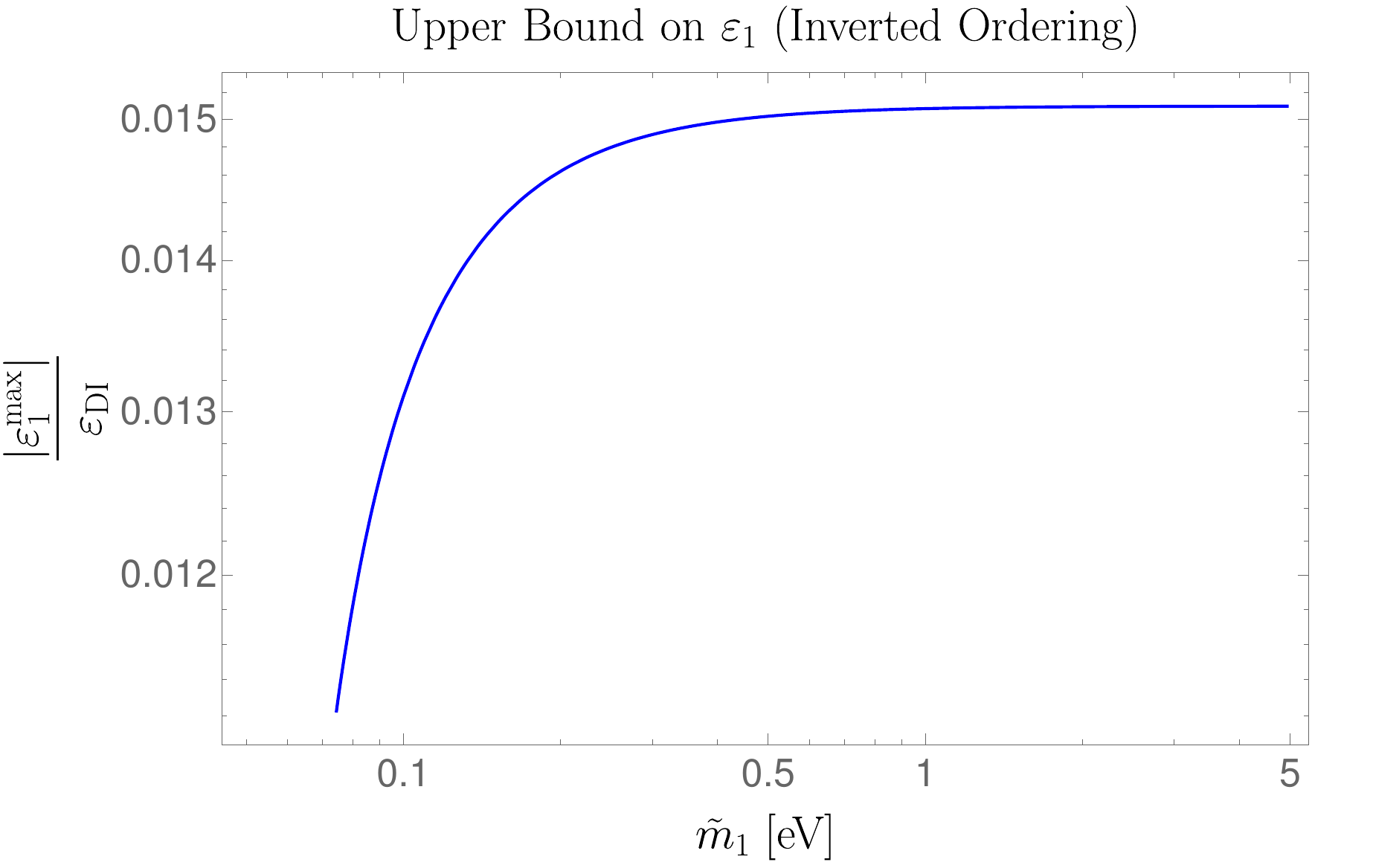}
\end{center}    
\caption{CP asymmetry upper bound $|\varepsilon_1^{\textrm{max}}|$ from $N_1$ decays in  two RHN scenario as a function of $\tilde{m}_1$ normalized by Davidson-Ibarra bound~\cite{Davidson:2002qv}. The left figure corresponds to the normal mass hierarchy case showing the analytical bound obtained from Eq.~\eqref{eq:ExactAnalytocalboundonepsilon1}. The right figure corresponds to the inverted neutrino mass hierarchy scenario showing the analytical bound obtained from Eq.~\eqref{eq:ExactAnalytocalboundonepsilon1InvertedHierarchy}. Also, it might be worthwhile to note the the plot for inverted hierarchy (right) starts  for much larger values of $\tilde{m}_1$ compared to the corresponding plot for normal hierarchy (left). This is because, as shown in Eq.~\eqref{eq:effectiveneutrinomasslowerbound}, $\tilde{m}_1$ is bounded below by $m_{\nu_2}$ (which becomes $m_{\nu_1}$ for inverted ordering), which is much larger in the inverted hierarchy case.}
\label{fig:CP_Violation_Upper_Bound}
\end{figure}

As mentioned earlier, once the neutrino mass ordering is fixed, $m_{\nu_2}$ and $m_{\nu_3}$ are fixed. For normal ordering, we have $m_{\nu_2}=\sqrt{\Delta m_{\textrm{sol}}^2}\approx 8.6\times 10^{-3}$ eV and $m_{\nu_3}=\sqrt{\Delta m_{\textrm{sol}}^2+\Delta m_{\textrm{atm}}^2}\approx$ 0.05 eV, where $\Delta m_{\textrm{sol}}^2\simeq{7.37\times10^{-5}}$ $\textrm{eV}^2$ and $\Delta m_{\textrm{atm}}^2\simeq{2.5\times10^{-3}}$ $\textrm{eV}^2$ are the solar and atmospheric neutrino mass parameters respectively~\cite{Capozzi:2025wyn}. Upon obtaining $m_{\nu_2}$ and $m_{\nu_3}$, the coefficients of the cubic polynomial $\Delta(k)$, which depend on $m_{\nu_2}$, $m_{\nu_3}$ and $\tilde{m}_1$ are exactly known up to $\tilde{m}_1$ and we can exactly determine the root $k_2(\tilde{m}_1)$. Using this information in eq.\eqref{eq:ExactAnalytocalboundonepsilon1}, we can evaluate the exact analytical upper bound on $|\varepsilon_1|$ as a function of $\tilde{m}_1$. The left plot in Figure \ref{fig:CP_Violation_Upper_Bound} shows the variation of $|\varepsilon_1^{\textrm{max}}|$ as a function of $\tilde{m}_1$ for normal ordering.
We find that $|\varepsilon_1^{\textrm{max}}(\tilde{m}_1)|$ is a monotonically increasing function of $\tilde{m}_1$ asymptotically saturating as $\tilde{m}_1\rightarrow\infty$. Therefore, the absolute global maximum of the CP violation $|\varepsilon_1|$ is reached in the $\tilde{m}_1\rightarrow\infty$ limit and is interestingly is less than the Davidson-Ibarra bound~\cite{Davidson:2002qv} by a factor of $\left(1-\frac{m_{\nu_2}}{m_{\nu_3}}\right)$. One heuristic way to understand this is that the number of parameters  in two RHN scenario is much less than the more standard three RHN scenario implying a tighter available parameter space for optimizing the parameters to maximize $|\varepsilon_1|$ and hence, a lower (and hence, stronger) upper bound on $|\varepsilon_1|$ than the Davidson-Ibarra bound in the standard three RHN scenario. We note that the global maximum of $|\varepsilon_1|$ that we obtained in the large $\tilde{m}_1$ regime coincides with the global upper bound of CP-asymmetry from $N_1$ decay in the scenario with two RHNs derived in \cite{Drees:2024hok}, which acts as an independant verification to our result. However, it is to be noted that in \cite{Drees:2024hok}, only a global upper bound on $|\varepsilon_1|$ was derived, which is valid in the large $\tilde{m}_1$ regime, in contrast to our explicit analytical form of $\tilde{m}_1$ dependent upper bound in Eq.~\eqref{eq:ExactAnalytocalboundonepsilon1}.\\


\subsection{Inverted hierarchy}\label{subsec:CPviolationInvertedOrdering}
In the case of inverted hierarchy, we reiterate that the results, mathematical steps and reasoning applied earlier are equally applicable and the exact analytical expression of the upper bound on CP asymmetry is obtained from Eq.~\eqref{eq:ExactAnalytocalboundonepsilon1} by relabeling the light neutrino mass eigenstates as $m_{\nu_1}\rightarrow m_{\nu_3}$, $m_{\nu_2}\rightarrow m_{\nu_1}$ and $m_{\nu_3}\rightarrow m_{\nu_2}$ so that we have
\begin{equation}
    \label{eq:ExactAnalytocalboundonepsilon1InvertedHierarchy}|{\varepsilon_1^{\textrm{max}}}|\equiv \varepsilon_{\textrm{DI}} \left(1-\frac{m_{\nu_1}}{m_{\nu_2}}\right)\left(1+\frac{m_{\nu_1}}{m_{\nu_2}}\right)\left(\frac{m_{\nu_2}}{\tilde{m}_1}\right)\sqrt{k_2 (\tilde{m}_1)} \hspace{2pt}.
\end{equation}
In this case, we have
we have $m_{\nu_1}=\sqrt{\Delta m_{\textrm{atm}}^2}\approx 0.0492$ eV and $m_{\nu_2}=\sqrt{{\Delta m_{\textrm{atm}}^2}+\Delta m_{\textrm{sol}}^2}\approx 0.05 $ eV, using which we find $|{\varepsilon_1^{\textrm{max}}}|$ , with the result shown in the right plot of Figure \ref{fig:CP_Violation_Upper_Bound}. Similar to normal hierarchy case, we find that $|\varepsilon_1^{\textrm{max}}|$ is a monotonically increasing function of $\tilde{m}_1$ asymptotically saturating as $\tilde{m}_1\rightarrow\infty$ and the global maximum of $|\varepsilon_1|$ is reached in the $\tilde{m}_1\rightarrow\infty$ limit. The global maximum is less than the Davidson-Ibarra bound~\cite{Davidson:2002qv}, similar to the normal hierarchy scenario, but in this case by a factor of $\left(1-\frac{m_{\nu_1}}{m_{\nu_2}}\right)$ . However, for the inverted hierarchy, the deviation due to this factor is much larger than normal hierarchy case and the global maximum is almost two orders of magnitude smaller than the Davidson-Ibarra bound. 

\section{A new lower bound on $m_{N_1}$}\label{sec: A new lower bound on lightest RHN mass}
\begin{figure}[t]
\begin{center}
\includegraphics[width=0.8\linewidth]{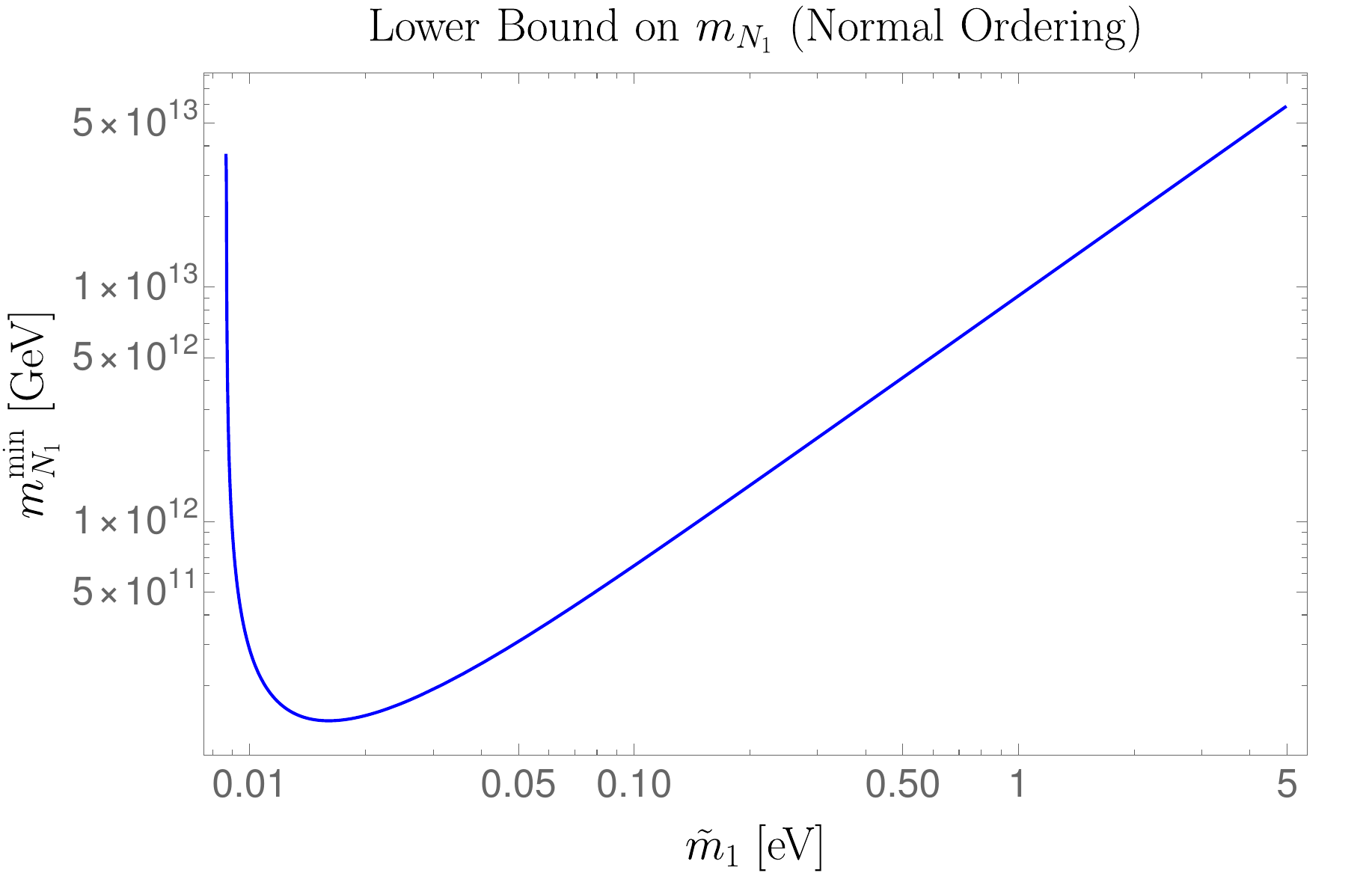}
\end{center}    
\caption{The estimated lower bound $m_{N_1}^\textrm{min}$  as a function of effective neutrino mass $\tilde{m}_1$ required for successful vanilla leptogenesis  in the two RHN scenario in the strong hierarchy limit in the normal neutrino mass hierarchy scenario assuming zero initial abundance of $N_1$. In principle, leptogenesis can be realized throughout the region above the blue curve, subject to perturbativity and washout constraints. We confine ourself only within the region where $m_{N_1}^\textrm{min}<10^{14}$ GeV.}
\label{fig:lowerboundmN1}
\end{figure}

As mentioned earlier, in order to generate the observed baryon asymmetry, $|\varepsilon_1|$ must satisfy Eq.~\eqref{eq:CP violationObservedBaryonAsymmetry}. But simultaneously, it should also be bounded above by $|\varepsilon_1^{\textrm{max}}|$ from Eq.~\eqref{eq:ExactAnalytocalboundonepsilon1}. Therefore, we have 
\begin{equation}
    |\varepsilon_1|\simeq\frac{6.2\times10^{-8}}{\kappa_f(\tilde{m}_1)}\leq \varepsilon_{DI} \left(1-\frac{m_{\nu_2}}{m_{\nu_3}}\right)\left(1+\frac{m_{\nu_2}}{m_{\nu_3}}\right)\left(\frac{m_{\nu_3}}{\tilde{m}_1}\right)\sqrt{k_2 (\tilde{m}_1)}
\end{equation}
rearranging which we get a lower bound on $m_{N_1}$
\begin{equation}
\label{eq:LowerBoundmN1}
    m_{N_1}\gtrsim \frac{6.2\times10^{-8}}{\kappa_f(\tilde{m}_1)} \frac{16\pi v^2}{3 m_{\nu_3}\left(1-\frac{m_{\nu_2}}{m_{\nu_3}}\right)\left(1+\frac{m_{\nu_2}}{m_{\nu_3}}\right)\left(\frac{m_{\nu_3}}{\tilde{m}_1}\right)\sqrt{k_2 (\tilde{m}_1)}} \equiv m_{N_1}^{\textrm{min}}(\tilde{m}_1)
\end{equation}
where we have used Eq.~\eqref{eq:DIbound} for $\varepsilon_{\textrm{DI}}$. As long as $m_{N_1}\ll 10^{14}$ GeV, we can approximate $\kappa_f$ by the semianalytical expression given in  Eq.~\eqref{eq:kappaonm1tilde}~\cite{Giudice:2003jh}.
For $m_{N_1}\gtrsim 10^{14}$ GeV, the situation becomes more complicated as $\kappa_f$ depends on both $\tilde{m}_1$ and $m_{N_1}  $, and there was no simple semianalytical form for $\kappa_f$ proposed in ~\cite{Buchmuller:2002rq,Buchmuller:2004nz,Giudice:2003jh} for fitting the numerical solution obtained by solving the Boltzmann equations in this regime.
So, we will only confine ourselves within the region where the estimated lower bound $m_{N_1}^{\textrm{min}}$ is less than $10^{14}$ GeV.   The lower bound on $m_{N_1}$ as a function of $\tilde{m}_1$ is shown in Figure \ref{fig:lowerboundmN1} for normal hierarchy assuming zero initial abundance of $N_1$. For inverted hierarchy, for almost the entire permissible range of $\tilde{m}_1$ , our rough estimate suggests that $m_{N_1}^\textrm{min}\geq 10^{14}$ GeV and therefore requires careful numerical analysis for reasons mentioned earlier. We do not dive further into this in the context of our work although a dedicated future study on this would be worthwhile. 
We find that the lower bound on $m_{N_1}$ as a function of effective neutrino mass $\tilde{m}_1$ required for vanilla leptogenesis in the 2 RHN scenario is much larger, and therefore stronger, than the lower bound on $m_{N_1}$ required for leptogenesis in the more standard 3 RHN scenario obtained in \cite{Buchmuller:2004nz,Giudice:2003jh}. 

\section{Conclusions}\label{sec: Conclusions}

We have derived an exact analytical expression of the upper bound on the magnitude of CP-asymmetry generated from the decay of the lightest heavy RHN $N_1$ in the type-I seesaw with two RHNs for standard vanilla leptogenesis in the hierarchical limit $m_{N_1}\ll m_{N_2}$ as a function of {effective neutrino mass} $\tilde{m}_1$. For standard vanilla leptogenesis, the final lepton and baryon asymmetry is determined by two parameters $\tilde{m}_1$ and $m_{N_1}$. We find that at every point in this two dimensional $\tilde{m}_1-m_{N_1}$ parameter space, the $R$ matrix can be uniquely determined for the 2 RHN scenario. The $R$ matrix can be parameterized by one complex parameter, or equivalently two real parameters. At every point in the $\tilde{m}_1-m_{N_1}$ parameter space, leptogenesis gives two constraints in terms of the R matrix elements- one coming from expressing $\tilde{m}_1$ and the other coming from expressing the CP violation $\varepsilon_1$. Both of these conditions are expressed in terms of $R$ matrix elements, such that the two real parameters of the $R$ matrix are determined. Eliminating one of the two real parameters of the $R$ matrix, we get a quartic equation in the other parameter. The requirement to get a physically admissible $R$ matrix compatible with leptogenesis imposes constraints on the nature of the roots of the quartic equation, from which we can derive an exact upper bound on $|\varepsilon_1|$ as a function of $\tilde{m}_1$. 

We find that the upper bound $|\varepsilon_1^{\textrm{max}}|$, given by Eq.~\eqref{eq:ExactAnalytocalboundonepsilon1} (Eq.\eqref{eq:ExactAnalytocalboundonepsilon1InvertedHierarchy} for inverted hierarchy), is a monotonically increasing function of $\tilde{m}_1$, asymptotically saturating as $\tilde{m}_1\rightarrow\infty$, implying that the global maximum of $|\varepsilon_1^{\textrm{max}}|$ is reached as $\tilde{m}_1\rightarrow\infty$. Interestingly, we find that the global maximum of CP violation in this case is less than the Davidson-Ibarra bound~\cite{Davidson:2002qv} by a factor of $\left(1-\frac{m_{\nu_2}}{m_{\nu_3}}\right)$ and $\left(1-\frac{m_{\nu_1}}{m_{\nu_2}}\right)$ in the normal and inverted hierarchy scenario respectively, implying a stronger bound on the CP asymmetry compared to the Davidson-Ibarra bound. For the inverted hierarchy, our exact global bound on the CP asymmetry is almost two orders of magnitude more stringent than the Davidson-Ibarra bound. 

It might be worthwhile to note that the global upper bound on $|\varepsilon_1|$ in the large $\tilde{m}_1$ limit obtained from our analytical expression coincides with the global upper bound on $|\varepsilon_1|$ for the scenario with two RHNs derived in \cite{Drees:2024hok}. However, as mentioned earlier, in \cite{Drees:2024hok}, only a global upper bound on $|\varepsilon_1|$ was derived, which is valid in the large $\tilde{m}_1$, in contrast to our explicit analytical bound valid for general $\tilde{m}_1$. Also, 
we compare the exact analytical upper bound on $|\varepsilon_1|$ as a function of $\tilde{m}_1$, that we derived, with the approximate analytical bound given in \cite{Ibarra:2003up}, where $\mathcal{O}\left(\frac{m_{\nu_2}}{m_{\nu_3}}\right)$ terms were ignored in the approximation. We find that our upper bound is  stronger. We also compared our exact analytical expression with the analytical expressions present in \cite{Guo:2006qa, DiBari:2005st} where we find numerical differences with our bound being more restrictive.

Finally, after deriving the exact analytical expression of the upper bound on $|\varepsilon_1|$, we use that to estimate the lower bound on $m_{N_1}$ as a function of $\tilde{m}_1$ that is compatible with standard vanilla leptogenesis for the two RHN scenario in the strong hierarchy limit. We present our results for normal neutrino mass hierarchy. We find that the lower bound on $m_{N_1}$ required for leptogenesis in the two RHN scenario is much larger, and therefore stronger, than the lower bound on $m_{N_1}$ required for leptogenesis in the more standard three RHN scenario derived in \cite{Buchmuller:2004nz,Giudice:2003jh}. This is in agreement with other works that explored leptogenesis in two RHN scenario~\cite{Okada:2025daq,Xing:2020ald,Bhattacharya:2006aw}. For inverted neutrino mass hierarchy, our rough estimate suggests that for most range of $\tilde{m}_1$, the lower bound on $m_{N_1}$ required for vanilla leptogenesis in the two RHN scenario is greater than $10^{14}$ GeV, implying that the semianalytical approximation of efficiency factor $\kappa_f$  used in Eq.~\eqref{eq:kappaonm1tilde} is no longer accurate and has to be obtained from the numerical solutions of the Boltzmann equations~\cite{Buchmuller:2002rq,Buchmuller:2004nz,Giudice:2003jh}. We do not attempt to dive further into this issue although a future investigation on this would be worthwhile.

\section*{ACKNOWLEDGEMENTS}
I am grateful to Brian Batell, Arnab Dasgupta, Akshay Ghalsasi and Alejandro Ibarra for the discussions and insights which have been very helpful to me. I also thank Brian Batell for providing feedback on the manuscript.
\appendix

\section{Some mathematical prerequisites}
\subsection{Nature of roots of a quartic equation}

\label{sec:Nature_of_Roots}

Let's consider a generic quartic equation $ax^4 + b x^3 + cx^2 +dx +e=0$, with coefficients $a,b,c,d,e$ all real and $a\neq 0$. The following quantities will be important in determining the nature of its roots and therefore, the nature of the roots of the quartic equation Eq.~\eqref{eq:GeneralQuarticinR}.
\begin{itemize}
    \item Discriminant $\Delta$ defined as
    \begin{equation}
        \begin{split}
            \label{eq:GeneralDiscriminantQuartic}
        &\Delta\equiv256a^3e^3 -192a^2bde^2-128a^2c^2e^2 +144 a^2 c d^2 e -27 a^2 d^4 \\
        &+ 144ab^2 ce^2-6ab^2d^2e-80abc^2de+18abcd^3+16ac^4e\\
        &-4ac^3d^2-27b^4e^2+18b^3cde-4b^3d^3-4b^2c^3e+b^2c^2d^2
        \end{split}
    \end{equation}
    \item $P\equiv8ac-3b^2$
    \item $D\equiv 64a^3e-16a^2c^2+16ab^2c-16a^2bd-3b^4$
\end{itemize}
The conditions corresponding to the nature of the roots are as follows:
\begin{itemize}
    \item If $\Delta<0$, then the quartic equation has two real roots and two non-real complex roots. The pair of non-real complex roots are conjugate to each other.
    \item If $\Delta>0$, either all four roots are real and distinct or all four roots are complex non-real. The nature of roots are conclusively decided by $P$ and $D$ parameters. If $P<0$ and $D<0$, all four roots are real and distinct while if $P>0$ or $D>0$, then the roots are two pairs of complex conjugate non-real numbers..
    
\end{itemize}

\subsection{Nature of roots of a cubic equation}
\label{app:Nature_of_Roots_Cubic_Equation}
Let's consider a generic cubic polynomial $a_0x^3+a_1x^2+a_2x+a_3$, with $a_0\neq 0$. The nature of roots of this cubic polynomial is determined by its discriminant $\tilde{\Delta}$ which is given by
\begin{equation}
    \label{eq:DiscriminantCubic}
    \tilde{\Delta}= 18 a_0a_1a_2a_3-4a_1^3a_3+a_1^2a_2^2-4a_0a_2^3-27a_0^2a_3^2 \hspace{2pt}.
\end{equation}
The nature of the roots are solely determined by the sign of $\tilde{\Delta}$ as follows:
\begin{itemize}
    \item If $\tilde{\Delta}>0$, the cubic polynomial has three distinct real roots.
    \item If $\tilde{\Delta}<0$, the cubic polynomial has one real and a pair of non-real complex conjugate roots.
\end{itemize}

\subsection{Descartes' Rule of Signs}
\label{app:RuleofSigns}

Descartes' rule of signs helps us to determine the number of positive and negative real roots of a polynomial. The rule is as follows.
Let $P(x)$ be a polynomial with real coefficients. The number of positive real roots of $P(x)$ is equal to the number of sign changes in the coefficients of $P(x)$ or less than that by an even number, while the number of negative real roots of $P(x)$ is the number of sign changes in the coefficients of $P(-x)$ or less than that by an even number.

\subsection{Intermediate value theorem}
\label{app:Bolzano_Thm}
The intermediate value theorem states that if $f(x)$ is a continuous function in the closed interval $[a,b]$ and $c\in[f(a),f(b)]$, then there exists at least one number $x_0$ in the closed interval $[a,b]$ such that $f(x_0)=c$.

The special case with $c=0$ is also known as Bolzano's theorem.

For the quartic polynomial in Eq.~\eqref{eq:GeneralQuarticinR}, since the coefficient of $r^4$ is positive, 
the quartic polynomial approaches $+\infty$ as $r$ approaches $-\infty$, implying that given a large positive number, let's say $\tilde{\eta}$, however arbitrarily large, we can always find a sufficiently large negative number, let's say $-\frac{1}{\epsilon}$ such that $f\left(-\frac{1}{\epsilon}\right)=\tilde{\eta}>0$. Now, if we prove that there exists a root negative real root $r'$ of $f'(r)$ such that $f(r')<0$, then let's see how Bolzano theorem implies existence of a negative real root of $f(r)$.\\
First of all, the quartic polynomial $f(r)$ is continuous everywhere and therefore in the closed interval $[-\frac{1}{\epsilon},r']$. Secondly, $0$ lies inside the closed interval $[f(r'),f\left(-\frac{1}{\epsilon}\right)]$ as $f(r')<0$ and $f\left(-\frac{1}{\epsilon}\right)>0$. Therefore, all the conditions of Bolzano's theorem are satisfied, implying that there exists an $r_0$ in the interval $[r',-\frac{1}{\epsilon}]$ such that $f(r_0)=0$. And since $r_0\leq r'<0$, it implies that the quartic polynomial has a negative real root $r_0$.

 \subsection{Resultant and discriminant }
 \label{app:Resultant}
 Given a polynomial 
 \begin{equation}
     \label{eq:px}
     p(x)=a_nx^n+a_{n-1}x^{n-1}+...+a_1x+a_0
 \end{equation}
 of degree $n$ with roots $\alpha_1,\hspace{2pt}\alpha_2,..,\alpha_n$
 and a polynomial
 \begin{equation}
     \label{eq:qx}
     q(x)=b_mx^m+b_{m-1}x^{m-1}+...+b_1x+b_0 \hspace{2pt},
 \end{equation}
 of degree $m$ with roots $\beta_1,\beta_2,..,\beta_m$, the resultant $\rho(p,q)$, also called the eliminant, is defined by~\cite{TrottM}
\begin{equation}
    \label{eq:Resultant}
    \rho(p,q)=a_n^m b_m^n \prod_{i-1}^n \prod_{j=1}^m \left(\alpha_i-\beta_j\right) \hspace{2pt}.
\end{equation}
 So, it can be seen from eq.\eqref{eq:Resultant} that if there is a common root shared by $p(x)$ and $q(x)$, the resultant $\rho(p,q)$ vanishes.

 The discriminant $\bar{\Delta}$ of any arbitrary polynomial $f(x)$
 \begin{equation}
     f(x)=c_k x^k+c_{k-1}x^{k-1}+c_{k-2}x^{k-2}+..+c_1x+c_0
 \end{equation}
 can be related to the resultant of $f(x)$ and its derivative $f'(x)$ as~\cite{DiscriminantandResultant}
 \begin{equation}
     \label{eq:DiscriminantResultant}
     \bar{\Delta}(f)=\frac{{(-1)}^{k(k-1)/2}\rho(f,f')}{c_k} \hspace{2pt}.
 \end{equation}

 If the polynomial $f$ and its derivative $f'$ share a root, then $\rho(f,f')=0$. Then it follows from eq.\eqref{eq:DiscriminantResultant} that the discriminant $\bar{\Delta}(f)$ vanishes.

\section{Detailed analysis of nature of roots in the 2 RHN scenario}\label{app:DetailedAnalysis}
Here, we present the detailed mathematical analysis of the nature of the roots of the quartic equation Eq.~\eqref{eq:GeneralQuarticinR}. We reiterate that we have implicitly assumed normal ordering to arrive at Eq.~\eqref{eq:GeneralQuarticinR} and we continue to do so here. The mathematical steps presented here will remain valid for inverted ordering as well subject to the relabeling $m_{\nu_1}\rightarrow m_{\nu_3}$, $m_{\nu_2}\rightarrow m_{\nu_1}$ and $m_{\nu_3}\rightarrow m_{\nu_2}$. Also, the quartic equation in Eq.~\eqref{eq:GeneralQuarticinR} will change accordingly.

To analyze the nature of roots in the quartic polynomial in $r$ in Eq.~\eqref{eq:GeneralQuarticinR}, we have to look at its discriminant $\Delta$, which, for a general quartic equation, is defined in Eq.~\eqref{eq:GeneralDiscriminantQuartic}.
The discriminant $\Delta$ of the quartic polynomial in $r$ in Eq.~\eqref{eq:GeneralQuarticinR} is 
\begin{flalign}
\begin{aligned}
    \label{eq:DiscriminantQuarticinR}
    &\Delta=\frac{{\left({m_{\nu_3}^2}-{m_{\nu_2}^2}\right)}^2}{{m_{\nu_3}^{12}}} \left[ 4{\left({m_{\nu_3}^2}-{m_{\nu_2}^2}\right)}^4 k^3 \right.\\
    &\left.+{\left({m_{\nu_3}^2}-{m_{\nu_2}^2}\right)}^2\left\{{\tilde{m}_1}^4+{m_{\nu_2}^4}+{m_{\nu_3}^4}-10{m_{\nu_2}^2}{m_{\nu_3}^2}\right. \left.-10{\tilde{m}_1}^2\left({m_{\nu_2}^2}+{m_{\nu_3}^2}\right)\right\}k^2   \right.\\
    &\left.-\left\{2{\tilde{m}_1}^6\left({m_{\nu_2}^2}+{m_{\nu_3}^2}\right)-{\tilde{m}_1}^4\left(8{m_{\nu_2}^4}-4{m_{\nu_2}^2}{m_{\nu_3}^2}+8{m_{\nu_3}^4}\right)\right.\right.\\
 &\left.\left.+2{\tilde{m}_1}^2\left({m_{\nu_2}^6}+2{m_{\nu_2}^4}{m_{\nu_3}^2}+2{m_{\nu_2}^2}{m_{\nu_3}^4}+
    {m_{\nu_3}^6}\right)\right.\right.\\
    &\left.\left.+2{m_{\nu_2}^2}{m_{\nu_3}^2}\left({m_{\nu_2}^4}-4{m_{\nu_2}^2}{m_{\nu_3}^2}+{m_{\nu_3}^4}\right)\right\}k+{\left({\tilde{m}_1}^2-{m_{\nu_2}^2}\right)}^2{\left({\tilde{m}_1}^2-{m_{\nu_3}^2}\right)}^2 \right] \hspace{5pt},
\end{aligned}    
\end{flalign}
where $k$ is defined in Eq.~\eqref{eq:DefinitionK}.
As seen in Eq.~\eqref{eq:DiscriminantQuarticinR}, the discriminant $\Delta$ can be thought of as a cubic polynomial in $k$. Therefore, from now on, we will denote the discriminant in Eq.~\eqref{eq:DiscriminantQuarticinR} as $\Delta(k)$, to emphasize this feature.

As elaborated on in Appendix \ref{sec:Nature_of_Roots}, the nature of roots of a general quartic equation depends on the sign of $\Delta$.
In order to find the sign of $\Delta(k)$, we will find the nature of roots of the cubic polynomial $\Delta(k)$, which in turn, depends on its discriminant $\tilde{\Delta}$. 

The discriminant of a general cubic polynomial is given in eq.\eqref{eq:DiscriminantCubic}, from which we can compute the discriminant $\tilde{\Delta}$ of the cubic polynomial $\Delta(k)$, which is
\begin{flalign}
\label{eq:Discriminantofdiscriminant}
\begin{aligned}
&\tilde{\Delta}= \frac{16 {\tilde{m}_1}^2 {m_{\nu_2}^2}}{m_{\nu_3}^{46}} {\left({m_{\nu_3}^2}-{m_{\nu_2}^2}\right)}^{12}\\
&{\left\{{\tilde{m}_1}^6+3{\tilde{m}_1}^4\left({m_{\nu_2}^2}+{m_{\nu_3}^2}\right)+{\left({m_{\nu_2}^2}+{m_{\nu_3}^2}\right)}^3+3{\tilde{m}_1}^2\left({m_{\nu_2}^4}-7{m_{\nu_2}^2}{m_{\nu_3}^2}+{m_{\nu_3}^4}\right)\right\}}^3  \hspace{1pt}.  
\end{aligned}
\end{flalign}
 In order to know the nature of roots of the cubic polynomial $\Delta(k)$, we need to know the sign of $\tilde{\Delta}$. 
For normal ordering, $m_{\nu_2}=\Delta m_{\textrm{sol}}$, $m_{\nu_3}=\sqrt{\Delta m_{\textrm{sol}}^2+\Delta m_{\textrm{atm}}^2}$, for which it can be seen that $\tilde{\Delta}>0$, implying that all three roots of the cubic polynomial $\Delta(k)$ are real and distinct. Let those roots be $k_1,\hspace{2pt}k_2$ and $k_3$ with $k_1>k_2>k_3$.

Since the common factor of the cubic polynomial $\Delta(k)$, which is $\frac{{\left({m_{\nu_3}^2}-{m_{\nu_2}^2}\right)}^2}{{m_{\nu_3}^{12}}}$, is always positive, we can factor it out and focus on the remaining polynomial. The roots of the remaining polynomial remain  $k_1,k_2,k_3$ and so do the conclusions regarding the sign of $\Delta(k)$.  Since the coefficient of $k^3$, which is $4{\left({m_{\nu_3}^2}-{m_{\nu_2}^2}\right)}^4$, is positive, we have $\Delta(k)>0$ for $k_3<k<k_2$ and $k>k_1$ while we have $\Delta(k)<0$ for $k<k_3$ and $k_2<k<k_1$.
Also, since $k$ is square of a real number (c.f. Eq.~\eqref{eq:DefinitionK}), we always have $k\geq 0$. Therefore, we need to determine the sign of $\Delta(k)$ in the regime $k\geq0$, which will be dictated by the number of positive roots of $\Delta(k)$. We use Descartes' rule of signs to determine that.

For the polynomial $\Delta(k)$ without the overall common factor $\frac{{\left({m_{\nu_3}^2}-{m_{\nu_2}^2}\right)}^2}{{m_{\nu_3}^{12}}}$, the coefficient of $k^3$, which is  $4{\left({m_{\nu_3}^2}-{m_{\nu_2}^2}\right)}^4$, and the constant term, which is ${\left({\tilde{m}_1}^2-{m_{\nu_2}^2}\right)}^2{\left({\tilde{m}_1}^2-{m_{\nu_3}^2}\right)}^2$, are both positive. The signs of coefficients of $k^2$ and $k$, which can be read off from Eq.~\eqref{eq:DiscriminantQuarticinR}, are positive or negative depending on the value of ${\tilde{m}_1}^2$. The coefficients of $k^2$ and $k$ can be thought of respectively as quadratic and cubic polynomials of ${\tilde{m}_1}^2$, and the domains where they are positive and negative can be determined by analysis of roots of these polynomials, keeping in mind that ${\tilde{m}_1}^2>0$. Table \ref{table:1} shows the signs of the coefficients of $k$ and $k^2$ for different ranges of ${\tilde{m}_1}^2$.

We use Descartes' rule of signs to determine the number of positive and negative real roots of $\Delta(k)$ for different ranges of values of ${\tilde{m}_1}^2$. We briefly present the arguments below.
\begin{itemize}
    \item $0<{\tilde{m}_1}^2<1.72\times10^{-4} \textrm{ eV}^2$: Here, the coefficient of $k^2$ is positive and coefficient of $k$ is negative in $\Delta(k)$. So there are two sign changes in $\Delta(k):$ positive to negative as we go from the coefficient of $k^2$ to coefficient of $k$ and negative to positive as we go from coefficient of $k$ to the constant term. So we have two or no positive real roots. In $\Delta(-k)$, the coefficient of $k^3$ is negative but all other coefficients are positive. So we have only one sign change as we go from the coefficient of $k^3$ to that of $k^2$. So we have only one negative real root. Since we know that all the three roots of the polynomial $\Delta(k)$ must be real and distinct, the number of positive real roots must be two. So we have two positive real roots and one negative real root.
    \item $1.72\times10^{-4} \textrm{ eV}^2<{\tilde{m}_1}^2<7.95\times10^{-4} \textrm{ eV}^2$: Both the coefficients of $k^2$ and $k$ in $\Delta(k)$ negative. So, there are two sign changes: one from $k^3$ to $k^2$ and the other from $k$ to constant term, implying either two or no positive real roots. In $\Delta(-k)$, coefficients of $k^3$ and $k^2$ are negative while the coefficients of $k$ and the constant term are positive. So there is one sign change from $k^2$ to $k$, implying one negative real root. Because there are three distinct real roots, the number of positive real roots must be two. So we have two positive and one negative real root.
    \item Using similar reasoning, we see that in the remaining ranges of ${\tilde{m}_1}^2$, there are two positive and one negative real root of $\Delta(k)$.
\end{itemize}
\tiny
\begin{center}
\begin{table}
\begin{tabular}{||c |c |c |c |c||} 
 \hline
 ${\tilde{m}_1}^2$ range (in $eV^2$) & $k$ coefficient & $k^2$ coefficient & No(s). of +ve real roots & No(s). of -ve real roots \\ [0.5ex] 
 \hline\hline
 $0<{\tilde{m}_1}^2<1.72\times10^{-4}$ & -ve & +ve & 2 & 1\\ 
 \hline
 $1.72\times10^{-4}<{\tilde{m}_1}^2<7.95\times10^{-4}$ & -ve & -ve & 2 & 1\\
 \hline
 $7.95\times10^{-4}<{\tilde{m}_1}^2<8.33\times10^{-3}$ & +ve & -ve & 2 & 1\\
 \hline
 $8.33\times10^{-3}<{\tilde{m}_1}^2<2.57\times10^{-2}$ & -ve & -ve & 2 & 1\\
 \hline
 ${\tilde{m}_1}^2>2.57\times10^{-2}$ & -ve & -ve & 2 & 1\\ [1ex] 
 \hline
\end{tabular}
 \caption{Nature of real roots for different ranges of values of ${\tilde{m}_1}^2$. Here,the notation '+ve' stands for positive and '-ve' stands for negative.}
 \label{table:1}
\end{table}
\end{center}
\normalsize

So, we come to the conclusion that the cubic polynomial $\Delta(k)$ has two positive roots and one negative root. So, we have $k_3<0<k_2<k_1$. Therefore, in the regime $k\geq0$, we have $\Delta(k)>0$ for $0\leq k<k_2$ and $k>k_1$ while $\Delta(k)<0$ for $k<2<k<k_1$.
 Therefore, from appendix \ref{sec:Nature_of_Roots}, we can conclude that for $k$ in the range $0\leq k<k_2$ and $k>k_1$, either all four roots of the quartic equation Eq.~\eqref{eq:GeneralQuarticinR} are real and distinct or all four roots are non-real while in the range $k_2<k<k_1$, two roots are real and distinct, while the other two roots are non-real. In the range $0\leq k<k_2$ and $k>k_1$, to draw conclusions about the nature of the roots, we need to evaluate the parameters $P$ and $D$, which are defined in appendix \ref{sec:Nature_of_Roots}, for the quartic equation Eq.~\eqref{eq:GeneralQuarticinR}. For convenience, from now on, we will refer to the regions $0\leq k<k_2$, $k_2<k<k_1$ and $k>k_1$ as regions I, II and III respectively.

\begin{figure}[t]
\includegraphics[width=0.8\textwidth]{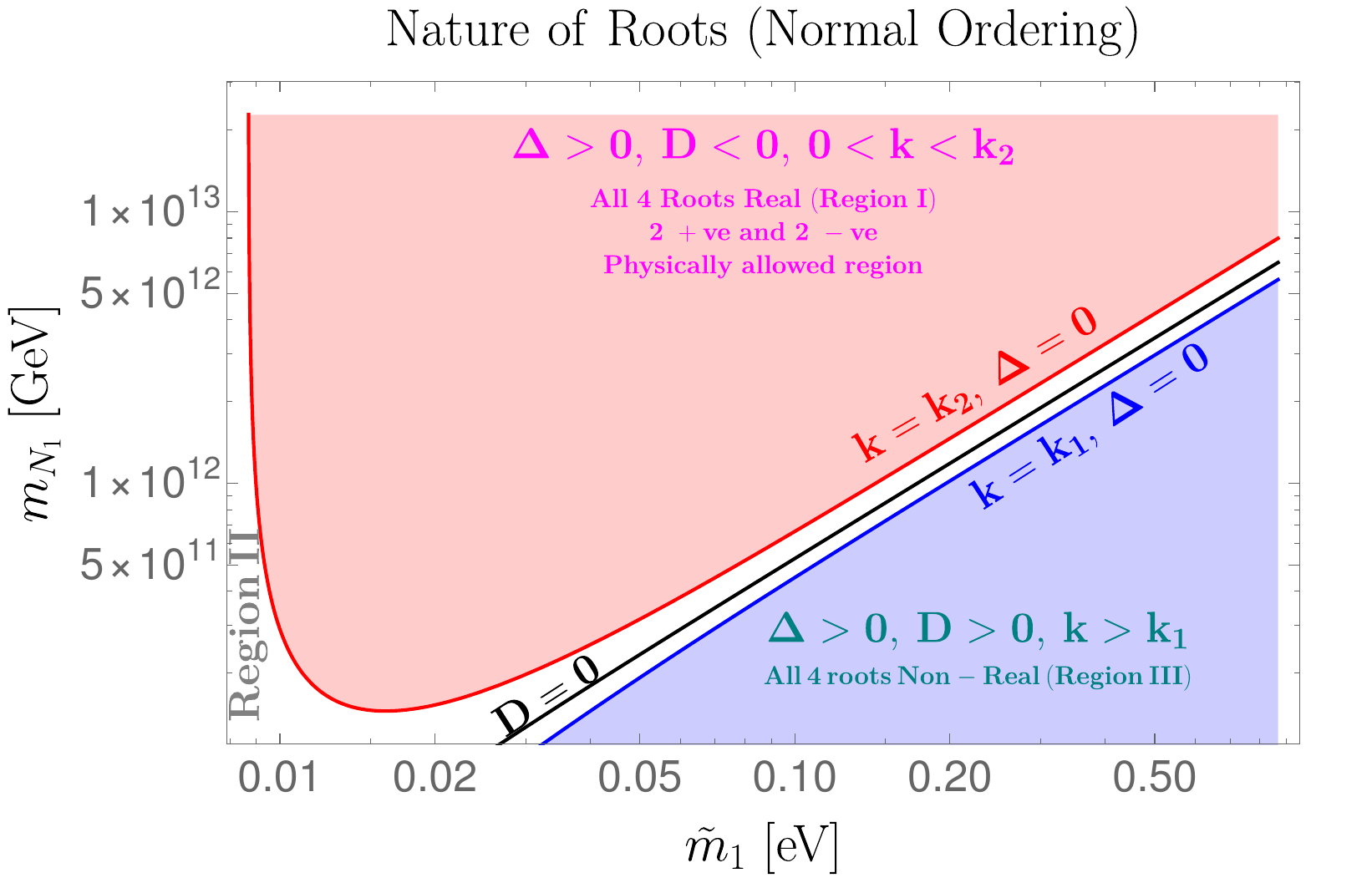}\\
\caption{Nature of the roots of Eq.~\eqref{eq:GeneralQuarticinR} for different regions in the $\tilde{m}_1-m_{N_1}$ parameter space for normal hierarchy of active neutrino masses. Regions I, II and III correspond to $0\leq k< k_2$, $k_2<k<k_1$ and $k>k_1$ respectively.  The red line corresponds to $k=k_2$. Above it is Region I in light red where $0<k<k_2$. Here, all four roots of Eq.~\eqref{eq:GeneralQuarticinR} are real: two positive and two negative. The positive real roots are the physically admissible solutions. The blue line corresponds to $k=k_1$. Below it is the region III in light blue where $k>k_1$. Here, all four roots of Eq.~\eqref{eq:GeneralQuarticinR} are non-real. The white region between the red and blue lines correspond to the region $k_2<k<k_1$ which is region II. Here, two roots are real (both negative, and hence unphysical) and the remaining two are non-real. The black line corresponds to $D=0$ , or equivalently Eq.~\eqref{eq:Dparameterzeroquarticm1tildemN1space}. Below it, $D>0$ and above it $D<0$.
The redline also corresponds to the lower bound on $m_{N_1}$ as derived in Eq.~\eqref{eq:LowerBoundmN1}.
}
\label{fig:Solution_nature_Normal_Ordering}
\end{figure}

For the quartic equation Eq.~\eqref{eq:GeneralQuarticinR}, the $P$ parameter is 
\begin{equation}
    \label{eq:PParameterQuartic}
    P=-\frac{{\left({m_{\nu_3}^2} -{m_{\nu_2}^2}\right)}^2 \left({\tilde{m}_1}^2 +{m_{\nu_2}^2} +{m_{\nu_3}^2}\right)}{{m_{\nu_3}^6}} \hspace{2pt},
\end{equation}
which, as evident from Eq.~\eqref{eq:PParameterQuartic}, is always negative. And the corresponding $D$ parameter is given by
\begin{equation}
  \label{eq:DParameterQuartic}
  D=-{\left({m_{\nu_3}^2} -{m_{\nu_2}^2}\right)}^4 \hspace{12pt}\frac{{m_{\nu_2}^2} {m_{\nu_3}^2} - k{\left({m_{\nu_3}^2} -{m_{\nu_2}^2}\right)}^2 + {\tilde{m}_1}^2\left({m_{\nu_2}^2}+{m_{\nu_3}^2}\right)}{{m_{\nu_3}^{12}}} \hspace{5pt}.
\end{equation}
The condition $D=0$ corresponds to 
\begin{equation}
    \label{eq:Dparameterzeroquartickspace}
    k= \frac{{m_{\nu_2}^2} {m_{\nu_3}^2} +{\tilde{m}_1}^2\left({m_{\nu_2}^2}+{m_{\nu_3}^2}\right)}{{\left({m_{\nu_3}^2} -{m_{\nu_2}^2}\right)}^2} \hspace{5pt},
\end{equation}
or equivalently 
\begin{equation}
    \label{eq:Dparameterzeroquarticm1tildemN1space}
    m_{N_1}= \frac{16\pi v^2  \tilde{m}_1|\varepsilon_1\left(\tilde{m}_1\right)|}{3\sqrt{{m_{\nu_2}^2} {m_{\nu_3}^2 }+{\tilde{m}_1}^2\left({m_{\nu_2}^2}+{m_{\nu_3}^2}\right)}} \hspace{5pt},
\end{equation}
with $|\varepsilon_1\left(\tilde{m}_1\right)|$ given by Eq.~\eqref{eq:CP violationObservedBaryonAsymmetry}. It can be shown that in region I, $D<0$, implying that all four roots of the quartic equation Eq.~\eqref{eq:GeneralQuarticinR} are real and distinct while in region III, $D>0$, implying that all four roots are non-real. Also, in region II, $\Delta(k)<0$, implying that two roots are real and distinct while the other two are non-real. The three regions I, II and III can be translated into the $\tilde{m}_1-m_{N_1}$ parameter space from the $k$ parameter space using Eq.~\eqref{eq:DefinitionK}\footnote{In Eq.~\eqref{eq:DefinitionK}, we use Eq.~\eqref{eq:CP violationObservedBaryonAsymmetry} to obtain $|\varepsilon_1|$, which implicitly assumes zero initial abundance of $N_1$ for $\kappa_f$ (c.f. Eq.~\eqref{eq:kappaonm1tilde}). However, the reasoning and analyses presented are independent of the explicit form of $\kappa_f$  and holds true regardless of the  $N_1$ initial abundance.}. Figure \ref{fig:Solution_nature_Normal_Ordering} shows the three regions I, II and III in the $\tilde{m}_1-m_{N_1}$ parameter space and the nature of roots in each region. The signs of $\Delta$, $P$ and $D$ in each of the three regions, and correspondingly, the nature of the roots, as concluded from Appendix \ref{app:DetailedAnalysis}, is summarized in Table \ref{table:2}.

\begin{center}
\begin{table}
\begin{tabular}{||c |c |c |c |c||} 
 \hline
 Region &$\Delta$ sign& $P$ sign &$D$ sign & Nature of roots \\  
 \hline\hline
 Region I & +ve & -ve & -ve & All 4 real roots\\ 
 \hline
 Region II & -ve & -ve & Both & 2 real, 2 non-real roots\\
 \hline
 Region III & +ve & -ve & +ve & All 4 nonreal roots\\
 \hline
\end{tabular}
 \caption{Signs of $\Delta$, $P$ , $D$ and nature of the roots in all the three regions. The symbols '+ve' and '-ve' stand for positive and negative respectively.}
 \label{table:2}
\end{table}
\end{center}

Now that we know that real solutions of Eq.~\eqref{eq:GeneralQuarticinR} are guranteed to exist in regions I and II, we need to determine if any of those real solutions are guranteed to be positive. This is simply because, as mentioned earlier, for the $R$-matrix in Eq.~\eqref{eq:RT} to be physically admissible and compatible with leptogenesis for a given solution $r$ of Eq.~\eqref{eq:GeneralQuarticinR}, we must have $r\geq 0$.

Next, we use Descartes' rule of signs to determine the possible number of positive and negative real solutions to Eq.~\eqref{eq:GeneralQuarticinR}.

Looking at Eq.~\eqref{eq:GeneralQuarticinR}, we see that the coefficients of $r^4,\hspace{5pt}r^3$ and the constant term 
are positive, while the coefficient of $r^2$
is negative. The coefficient of $r$, which is $ \frac{\tilde{m}_1m_{\nu_2}}{{m_{\nu_3}^2}}\left(1-\frac{{\tilde{m}_1}^2}{{m_{\nu_3}^2}}\right)$, is positive in the regime $\tilde{m}_1<m_{\nu_3}$ and negative in the regime $\tilde{m}_1>m_{\nu_3}$. So let's look at the nature of roots in either of these regimes i.e. in $\tilde{m}_1<m_{\nu_3}$ and $\tilde{m}_1>m_{\nu_3}$.

\begin{itemize}
    \item $\tilde{m}_1<m_{\nu_3}$: Here, the coefficients of $r^4, \hspace{2pt} r^3, \hspace{2pt} r$ and the constant term are positive, while the coefficient of $r^2$ is negative. So, we have two sign changes as we go from $r^4$ to the constant term: positive to negative from $r^3$ to $r^2$, and negative to positive from $r^2$ to $r$. Therefore, from Descartes' rule of signs, we have two or zero positive real roots.

    Also, for $r\rightarrow -r$, the coefficient of $r^4$ and the constant term are positive, while the coefficient of $r^3,\hspace{2pt}r^2$ and $r$ are negative, which means that there are two sign changes as we go from $r^4$ to the constant term: positive to negative from $r^4$ to $r^3$, and negative to positive from $r$ to the constant term. This implies, from the rule of signs, that there are two or zero negative real roots.

    \item $\tilde{m}_1<m_{\nu_3}$: Here, the coefficients of $r^4, \hspace{2pt} r^3$ and the constant term are positive, while the coefficients of $r^2$ and $r$  are negative. So, we have two sign changes as we go from $r^4$ to the constant term: positive to negative from $r^3$ to $r^2$, and negative to positive from $r$ to the constant term. Therefore, from Descartes' rule of signs, we have two or zero positive real roots.

    Also, for $r\rightarrow -r$, the coefficient of $r^4$, $r$ and the constant term are positive, while the coefficient of $r^3$ and $r^2$ are negative, which means that there are two sign changes as we go from $r^4$ to the constant term: positive to negative from $r^4$ to $r^3$, and negative to positive from $r^2$ to $r$. This implies, from the rule of signs, that there are two or zero negative real roots.
\end{itemize}

So, the conclusion is that the quartic polynomial equation in Eq.~\eqref{eq:GeneralQuarticinR} has two or zero positive real solutions and two or zero negative real solutions. Therefore, in region I, where all the four roots are guaranteed to be real and distinct, two of them must be positive and the other two must be negative. It is these two positive real solutions of $r$ that correspond to the $R$-matrices compatible with leptogenesis and hence, are the physically feasible solutions. Now, if we prove that in region II, where we have two real solutions, both the real solutions are negative, we can conclude the only region physically compatible with leptogenesis is $0<k<k_2$, and hence, the completes the reasoning reasoning that we presented in section \ref{sec: A new upper bound in CP violation} around Eq.~\eqref{eq:permissiblerangeofk} to derive the upper bound of $|\varepsilon_1|$ given by Eq.~\eqref{eq:ExactAnalytocalboundonepsilon1}.

So, finally, we move on to the last strand of our argument, that both the roots in Region II ($k_2<k<k_1$) are negative.

Since we know there are two or zero positive real roots and  two or zero negative real roots to the polynomial in Eq.~\eqref{eq:GeneralQuarticinR}, it implies that in region II, both the real roots are either positive or negative. So, it suffices to prove that if at least one of the real roots is negative, then both are negative. So, our strategy to prove is the following:

If the derivative $f'(r)$ of the quartic polynomial $f(r)$ in Eq.~\eqref{eq:GeneralQuarticinR} has one negative root $r'$ such that $f(r')<0$, then from Intermediate  value theorem/Bolzano theorem\footnote{Please refer to Appendix \ref{app:Bolzano_Thm} for details.}, we know that $f(r)$ has a root between $-\infty$ and $r'(<0)$, since we know that $\lim_{r\rightarrow -\infty} f(r)>0$, as the coefficient of $r^4$ in $f(r)$ is positive. So, we will show that the derivative polynomial $f'(r)$ indeed has a root $r'<0$ such that $f(r')< 0$.

The derivative polynomial $f'(r)$ is given by

\begin{flalign}
    \label{eq:2RHNbound16}
\begin{aligned}    
&f'(r)={\left(\frac{{m_{\nu_2}}^2}{{m_{\nu_3}}^2}-1\right)}^2 r^3 +3 \frac{\tilde{m}_1 m_{\nu_2}}{{m_{\nu_3}}^2}\left(1-\frac{{m_{\nu_2}}^2}{{m_{\nu_3}}^2}\right) r^2\\
&-\left\{1+\frac{{m_{\nu_2}}^2}{{m_{\nu_3}}^2} + \frac{{\tilde{m}_1}^2}{{m_{\nu_3}}^2} \left(1- 3 \frac{{m_{\nu_2}}^2}{{m_{\nu_3}}^2}\right) \right\} r
 + \frac{\tilde{m}_1 m_{\nu_2}}{{m_{\nu_3}}^2}\left(1-\frac{{\tilde{m}_1}^2}{{m_{\nu_3}}^2}\right) \hspace{5pt}.
\end{aligned} 
\end{flalign}

This is a cubic polynomial in $r$.
The discriminant of this cubic polynomial is

\begin{equation}
    \label{eq:2RHNbound17}
    \frac{4{\left({m_{\nu_3}}^2-{m_{\nu_2}}^2\right)}^2 \left\{\right\{ {\tilde{m}_1}^6 + 3{\tilde{m}_1}^4 \left({m_{\nu_2}}^2+{m_{\nu_3}}^2\right)+3{\tilde{m}_1}^2 \left({m_{\nu_3}}^4-7{m_{\nu_2}}^2 {m_{\nu_3}}^2 + {m_{\nu_2}}^4\right) +{\left({m_{\nu_2}}^2+{m_{\nu_3}}^2\right)}^3 \}}{{m_{\nu_3}}^{10}},
\end{equation}
which is positive, 
implying that $f'(r)$ has three distinct real positive roots, as inferred from Appendix \ref{app:Nature_of_Roots_Cubic_Equation}. We will show that out of these three roots, at least one is negative. We use Descartes' rule of signs to do that.

We divide our argument into two cases: ${\tilde{m}_1}<{m_{\nu_3}}$ and ${\tilde{m}_1}>{m_{\nu_3}}$.
\begin{itemize}
    \item ${\tilde{m}_1}<{m_{\nu_3}}$: In $f'(r)$, the coefficients of $r^3,\hspace{2pt}r^2$ and the constant term are positive, while coefficient of $r$ is negative. So we have two sign changes as we go from $r^3$ to the constant term: Positive to negative as we go from $r^2$ to $r$, and negative to positive as we go from $r$ to the constant term. So by Descartes' rule of signs, we have two or zero positive real roots.
    For $r\rightarrow -r$, the coefficient of $r^3$ is negative while the coefficient of $r^2, \hspace{2pt}r$ and the constant term is positive. So, there's one sign change- negative to positive as we go from $r^3$ to $r^2$. So by Descartes' rule of signs, we have one negative real root. Since we know that all three roots are real and distinct, we conclude that two of the real roots are positive and one is negative.
    \item ${\tilde{m}_1}>{m_{\nu_3}}$: In $f'(r)$, the coefficients of $r^3$ and $r^2$ are positive, while the coefficient of $r$ and the constant term are negative. So there's one sign change as we go from $r^3$ to $r^2$. So one real root is positive. For $r\rightarrow -r$, the coefficients of $r^3$ and the constant term is negative while the coefficients of $r^2$ and $r$ are positive. So as we go from $r^3$ to the constant term, there's a sign change of negative to positive from $r^3$ to $r^2$, and a sign change of positive to negative to $r$ to the constant term. So by Descartes' rule of signs, there are two or zero negative real roots. As all the three roots are real, we conclude that there are one positive and two negative distinct real roots.
\end{itemize}

So we see that the cubic polynomial $f'(r)$ has at least one negative real root (and at least one positive real root). 

Let the roots of $f'(r)$ be ${r_1}',\hspace{2pt}{r_2}'$ and ${r_3}'$,such that ${r_1}'>{r_2}'>{r_3}'$, with ${r_3}'<0$. 
We want to show that in region II, $f({r_3}')<0$. If we plot the curve $f({r_3}')=0$ in the ${\tilde{m}_1}-{m_{N_1}}$ parameter space, we see that the curve exactly overlaps with the blue curve, which corresponds to $k=k_1$ and $\Delta=0$, in Figure~\ref{fig:Solution_nature_Normal_Ordering}, which is also the boundary between Region II and Region III. Above this curve, which corresponds to region II (and also region I), we have $f({r_3}')<0$, implying the existence of one negative root, which is what we wanted to show. Also, the fact that the curve corresponding to $f({r_3}')=0$ exactly overlaps the blue curve  is not a coincidence. On the  blue curve, $f({r_3}')=0$, implying that ${r_3}'$ is a root of the polynomial $f(r)$ as well in addition to being the root of the polymial $f'(r)$. And it can be shown  that for any polynomial, when the polynomial shares at least one root with its derivative, the discriminant vanishes\footnote{Please see Appendix \ref{app:Resultant} for details.}. And this is exactly what happens on the blue line, where the discriminant $\Delta$ of the quartic polynomial vanishes.
\bibliographystyle{apsrev}
\bibstyle{apsrev}
\bibliography{ref.bib}
\end{document}